# Computationally Efficient Worst-Case Analysis of Flow-Controlled Networks With Network Calculus

Raffaele Zippo<sup>●</sup> and Giovanni Stea<sup>●</sup>

*Abstract*—Networks with hop-by-hop flow control occur in several contexts, from data centers to systems architectures (e.g., wormhole-routing networks on chip). A worst-case end-to-end delay in such networks can be computed using Network Calculus (NC), an algebraic theory where traffic and service guarantees are represented as curves in a Cartesian plane. NC uses transformation operations, e.g., the min-plus convolution, to model how the traffic profile changes with the traversal of network nodes. NC allows one to model flow-controlled systems, hence one can compute the end-to-end *service curve* describing the minimum service guaranteed to a flow traversing a tandem of flow-controlled nodes. However, while the algebraic expression of such an end-to-end service curve is quite compact, its computation is often intractable from an algorithmic standpoint: data structures tend to grow quickly to unfeasibly large sizes, making operations intractable, even with as few as three hops. In this paper, we propose computational and algebraic techniques to mitigate the above problem. We show that existing techniques (such as reduction to *compact domains*) cannot be used in this case, and propose an arsenal of solutions, which include methods to mitigate the data representation space explosion as well as computationally efficient algorithms for the min-plus convolution operation. We show that our solutions allow a significant speedup, enable analysis of previously unfeasible case studies, and - since they do not rely on any approximation - still provide exact results.

## I. INTRODUCTION

THE last few years have witnessed a surge in real-time networked applications, such as factory automation or collaborative robotics, within the Industry 4.0 paradigm, or self-driving / teleoperated cars. This has sparked a renewed industrial interest for methods that allow one to compute *performance bounds*, especially worst-case ones, since the above applications are clearly safety critical. Several systems supporting these applications communicate through networks with hop-by-hop window flow control. This occurs whenever a receiver has limited buffer space and wants to prevent the sender from overflowing it – because the loss of data is undesirable or too costly. This mechanism is used, for instance, in data center networks [1], [2], [3], or wormhole-routing networks on chip [4], [5]. Multi-hop networks with flow control have often been analyzed via classical queueing theory (see, e.g., [6]), that allows one to find probabilistic performance metrics using stochastic models of traffic and service. A worst-case analysis of such networks can be done via Network Calculus (NC). The latter is a theory for deterministic network evaluation, which dates back to the early 1990s, and it is mainly due to the work of Cruz [7], [8], Le Boudec and Thiran [9], and Chang [10]. Originally devised for the Internet, where it was used to engineer models of service [11], [12], [13], [14], it has found applications in several other contexts, from sensor networks [15] to avionic networks [16], [17], industrial networks [18], [19], [20], automotive systems [21] and systems architecture [22], [23]. Its main strength is that it allows one to compute worst-case delay bounds in systems with multi-hop traversal. It characterizes constraints on traffic arrivals (due to traffic shaping) and on minimum received service (due to scheduling) as *curves*, i.e., functions of time, and uses min-plus algebra to combine the above in order to compute bounds on the traffic at any point in a network traversal. More in detail, a network node is characterized by its *service curve*, a characteristic function that yields the worst-case response to an infinite burst (similarly to the transfer function in systems theory). When a flow traverses two nodes in tandem, its worst-case end-to-end service can be found by combining the service curves of the two nodes via their *min-plus convolution*. This operation yields a network-wide service curve, hence multi-node traversal can always be collapsed to single-node traversal via repeated convolutions. Most network nodes have simple service curves, called *rate-latency*, that can be represented by two segments: an initial horizontal segment – modeling the node's latency – followed by an infinite line whose slope is the node's rate. The convolution of two rate-latency curves is a rate-latency curve itself.

Window flow control can be modeled in NC. The algebraic operator that is required to do this is called *sub-additive closure* (SAC), which – as the name suggests – yields sub-additive curves. A flow-controlled node can thus be represented via an *equivalent* service curve, obtained via a SAC operation. Analysis of networks with hop-by-hop flow-control is made difficult by computational aspects. An equivalent service curve

Manuscript received 24 December 2021; revised 27 September 2022; accepted 31 January 2023. Date of publication 13 February 2023; date of current version 17 March 2023. This work was supported in part by the Italian Ministry of Education and Research (MIUR) in the framework of the FoReLab Project, Departments of Excellence; and in part by the University of Pisa, through Analisi di reti complesse: dalla teoria alle applicazioni under Grant PRA 2020. *(Corresponding author: Giovanni Stea.)*
Raffaele Zippo is with the Dipartimento di Ingegneria dell'Informazione, Università di Firenze, 50139 Firenze, Italy, and also with the Dipartimento di Ingegneria dell'Informazione, Università di Pisa, 56122 Pisa, Italy (e-mail: raffaele.zippo@ing.unipi.it).
Giovanni Stea is with the Dipartimento di Ingegneria dell'Informazione, Università di Pisa, 56122 Pisa, Italy (e-mail: giovanni.stea@unipi.it).

obtained via a SAC is unavoidably a staircase-based ultimately pseudo-periodic (UPP) function [24], [25], even when the node has a service curve as simple as a rate-latency one. UPP functions have an initial transient and a periodic part. Computing the worst-case delay of a flow traversing a tandem of flow-controlled hops requires one to compute an *end-to-end equivalent service curve* for the tandem first. However, this requires one to compute *nested* SACs, starting from per-node service curves. The algorithm to compute the SAC of a UPP curve is very complex, to the point that this analysis may be computationally unfeasible already with few hops (e.g., three). The SAC algorithm, in fact, requires performing a very large number of elementary convolutions. Work [26] introduces a different method, which dispenses with nested SACs. That method computes *per-node* equivalent service curves first, using SAC, and then computes the end-to-end equivalent service curve by *convolution* of per-node equivalent service curves. This second method yields an end-to-end equivalent service curve that lower bounds the one found with the former method. It is also less costly, since convolution of UPP curves is polynomial. However, chained convolutions of UPP curves may still be unfeasibly costly, due to a well-known phenomenon called *state explosion* ([27], [28]). It is observed therein that convolutions of UPP functions are often intractable, because the period of the result is tied to the least common multiple (lcm) of the periods of the operands. Thus, computing the end-to-end service curve of a tandem of $n$ nodes traversed by a flow by chaining $n-1$ convolutions – although *algebraically* simple – is often *computationally* intractable.

In [28], authors propose a method to mitigate this problem by observing that one may limit convolutions to a *compact* (i.e., finite) domain, without sacrificing accuracy. That finite domain is computed (at negligible cost) based on upper/lower approximations of UPP service curves and/or arrival curves. Using finite domains allows one to avoid the state explosion due to the lcm and the associated time complexity, making analysis faster – often by orders of magnitude. However, this method cannot be applied to our problem, since it relies on service curves being *super-additive*. The equivalent service curves that form the operands of the chained convolutions in our problem are instead *sub-additive* (having being computed via a SAC operation).

In this paper, we present computational and algebraic techniques to enable the analysis of flow-controlled networks on a larger scale, by reducing the number and time cost of the involved convolutions, often by orders of magnitude. We achieve this by performing two computationally simple tasks: first, minimizing the number of segments with which a UPP service curve is represented. This is particularly important, since – on one hand – the complexity of algorithms for basic min-plus operations (including convolution) depends on the number of segments of their operands, often in a superlinear way. On the other hand, the *number* of convolutions to be performed in a SAC depends on the number of segments of the operand. We show that *representation minimization* may reduce the number of segments by orders of magnitude. This makes operations generally faster – especially when several convolutions are chained together, and is of paramount importance to enable efficient SAC computation. Second, we prove algebraic properties of sub-additive functions that can be leveraged to drastically reduce the number of computations involved in the convolution. We prove that convolution of sub-additive UPP functions is in fact quite simple, unless the two operands intersect infinitely many times, and we prove a significantly faster algorithm for that case as well. We assess the gain in efficiency by measuring the achieved speedup of our findings on a desktop PC. As we show, the improvements are substantial, ranging from two-digit percentages to several orders of magnitude in most cases. Moreover, the speedups warranted by representation minimization and algebraic properties are cumulative. This not only makes computations more efficient: rather, it allows one to compute end-to-end service curves in cases where this was considered to be beyond the borders of tractability using the current NC methods and off-the-shelf hardware. Moreover, it allows one to compare the two methods discussed above, hence to benchmark the lower-bound approximation explained in [26] by efficiency and accuracy. Our findings in that respect are that the approximate method seems to be as accurate as the exact one, while considerably more efficient. Last, but not least, we remark that our techniques are *exact*, i.e., they do not entail any loss of accuracy in the end result. To the best of our knowledge, our results are novel and are not used in existing NC tools.

The rest of the paper is organized as follows: Section II introduces NC notation and basic results. We introduce the problem formally in Section III, and explain our techniques in Section IV. We report numerical evaluations in Section V. Section VI discusses the related works. Finally, Section VII concludes the paper and highlights directions for future work.

## II. NETWORK CALCULUS BASICS

We report here a necessarily concise introduction to NC, borrowing the notation used in [9], to which we refer the interested reader for more details.

A NC flow is represented as a wide-sense increasing and left-continuous cumulative function $R : \mathbb{R}_+ \to \mathbb{R}_+ \cup \{+\infty\}$. Function $R$ represents the number of bits of the flow observed in $[0, t[$. In particular, $R(0) = 0$.

Flows can be constrained by *arrival curves*. A wide-sense increasing function $\alpha$ is an *arrival curve* for a flow $A$ if:

$$\forall s \leq t, \quad A(t) - A(s) \leq \alpha(t-s).$$

For instance, a *leaky-bucket shaper*, with a *rate* $\rho$ and a *burst size* $\sigma$, enforces a concave affine arrival curve $\gamma_{\sigma,\rho}(t)$, defined as follows:

$$\gamma_{\sigma,\rho}(t) = \begin{cases} \sigma + \rho t, & \text{if } t > 0, \\ 0, & \text{otherwise.} \end{cases} \quad (1)$$

This means, among other things, that the long-term arrival rate of the flow cannot exceed $\rho$.

Let $A$ and $D$ be the functions that describe the same data flow at the input and output of a lossless network element (or *node*), respectively. If that node does not create data internally (which is often the case), causality requires that $A \geq D$.

We say that the node behavior can be modeled via a *service curve* $\beta$ if:

$$\forall t \geq 0, \quad D(t) \geq \inf_{0 \leq s \leq t} \{A(s) + \beta(t-s)\}. \quad (2)$$

In that case, the flow is guaranteed the (minimum) service curve $\beta$. The infimum on the right side of (2), as a function of $t$, is called the *(min-plus) convolution* of $A$ and $\beta$, and is denoted by $A \otimes \beta$. The alert reader can check that convolution is commutative and associative. Computing the above convolution entails sliding $\beta$ along $A$ and taking the *lower envelope* of the result (i.e., the infimum for each time instant).

Several network elements, such as delay elements, schedulers or links, can be modeled through service curves. A very frequent case is the one of *rate-latency* service curves, defined as:

$$\beta_{R,\theta}(t) = R[t - \theta]^+,$$

for some $\theta \geq 0$ (the latency) and $R > 0$ (the rate). Notation $(.)^+$ denotes $\max(.,0)$. For instance, a constant-rate server (e.g., a wired link) can be modeled as a rate-latency curve with zero latency.

A point of strength of NC is that service curves are *composable*: the end-to-end service curve of a tandem of nodes traversed by the same flow can be computed as the convolution of the service curves of each node.

For a flow that traverses a service curve (be it the one of a single node, or the end-to-end service curve of a tandem computed as discussed above), a tight *upper bound* on the delay can be computed by combining its arrival curve $\alpha$ and the service curve $\beta$ itself, as follows:

$$h(\alpha, \beta) = \sup_{t \geq 0} \{\inf\{d \geq 0 \mid \alpha(t-d) \leq \beta(t)\}\}. \quad (3)$$

The quantity $h(\alpha, \beta)$ is in fact the maximum horizontal distance between $\alpha$ and $\beta$. Therefore, computing the end-to-end service curve of a flow in a tandem traversal is the crucial step towards obtaining its worst-case delay bound.

In NC, the *sub-additive closure (SAC)* of a wide-sense increasing function $f$ is defined as:

$$\overline{f(t)} = \inf_{n \geq 0} \left\{ f^{(n)}(t) \right\}, \quad (4)$$

where $f^{(n)}$ denotes the $n$-fold self-convolution of $f$, i.e., $f^{(0)} = \delta_0$, $f^{(1)} = f$, and $f^{(n)} = \bigotimes_{i=1}^n f$ for $n \geq 1$. Function $\delta_0$ is an infinite step in $t = 0^+$, i.e., $\delta^{(0)}(0) = 0$ and $\forall t > 0, \delta^{(0)}(t) = +\infty$. Note that $\overline{f}$ is sub-additive, as the name suggests [9, Theorem 3.1.10]. The formal definition of sub-additivity is the following:

*Definition 1 (Sub-Additive Function):* $f$ is sub-additive if and only if $\forall u, s, f(u) + f(s) \geq f(u+s)$.

Moreover, given a function $f$ such that $f(0) = 0$, if $f$ is sub-additive then $\overline{f} = f$. Otherwise, it is $\overline{f} \leq f$. Convolution does preserve sub-additivity, as per the following property:

*Property 1 (Convolution of Sub-Additive Functions):* If $f$ and $g$ are sub-additive functions, so is $f \otimes g$ [9, Theorem 3.1.9].

Moreover, the SAC of a minimum is the convolution of the SACs of the operands, i.e.:

*Property 2 (SAC of a Minimum):* $\overline{f \wedge g} = \overline{f} \otimes \overline{g}$ [9, Theorem 3.1.11].

We spend a few words to clarify a relevant issue. The two theorems from [9] that we cite in the above two properties assume that functions $f$ and $g$ are wide-sense increasing. However, as already observed in [24], the proof of these theorems still holds even without that hypothesis.

### A. Computational Representation of NC Functions and Algorithms

NC computations can be implemented in software. In order to do so, one needs to provide computational representations of NC functions (e.g., a cumulative function of a flow or a service curve) and well-formed algorithms for its operations, e.g., minimum and convolution. We therefore describe a general data structure that represents NC functions, and the algorithms to compute the main NC operations used in this paper. We adopt the widely accepted approach described in [24] and [25].

To represent NC functions, we focus on the set $\mathcal{U}$ of ultimately pseudo-periodic (UPP), piecewise affine $\mathbb{Q}_+ \to \mathbb{Q} \cup \{+\infty, -\infty\}$ functions, as in [24]. It is shown therein that this class of functions is stable w.r.t. all min-plus operations,[1] while functions $\mathbb{R}_+ \to \mathbb{R} \cup \{+\infty, -\infty\}$ are not. An alternative class of functions with such stability is $\mathbb{N}_0 \to \mathbb{R} \cup \{+\infty, -\infty\}$, however this is only feasible for models where time is discrete. We remark that functions in $\mathcal{U}$ are not necessarily wide-sense increasing. While NC functions are usually assumed to be so, in order to implement min-plus operations it is sometimes useful to include non-monotonic functions as well. Similarly, functions in $\mathcal{U}$ can assume infinite values. This is also useful for algebraic manipulations, e.g., to express a function as a minimum of two or more functions.

A function in $\mathcal{U}$ has an *initial transient* of length $T \geq 0$, and a *period* of length $d > 0$ and height $c \in \mathbb{Q} \cup \{+\infty, -\infty\}$. Ultimately pseudo-periodic means that:

$$\forall t \geq T, f(t+d) = f(t) + c. \quad (5)$$

We denote with $\rho$ the *pseudo-periodic slope* of a function $f$, i.e., $\rho = \frac{c}{d}$. Functions in $\mathcal{U}$ are piecewise affine. This means that they can be represented as sequences of *points* and *segments*. The above terms are defined as follows:

*Definition 2 (Point):* We define a *point* as a tuple

$$p_i := (t_i, f(t_i)), \qquad i \in \{1, \ldots, n\}.$$

*Definition 3 (Segment):* We define a *segment* as a tuple

$$s_i := \left(t_i, t_{i+1}, f(t_i^+), f(t_{i+1}^-)\right), \qquad i \in \{1, \ldots, n\},$$

which describes $f$ in the open interval $]t_i, t_{i+1}[$ in which it is affine, i.e., for all $t \in ]t_i, t_{i+1}[$,

$$f(t) = f(t_i^+) + \frac{f(t_{i+1}^-) - f(t_i^+)}{t_{i+1} - t_i} \cdot (t - t_i) =: b + r \cdot (t - t_i)$$

---

[1] To be precise, the fact that an operation is well defined and stable in $\mathcal{U}$ may require additional properties, such as operands being *plain* or *ultimately plain*, as sufficient (but not necessary) conditions. We refer the reader to [24] for a thorough discussion. All the operations discussed in this paper are well-defined and stable in $\mathcal{U}$.

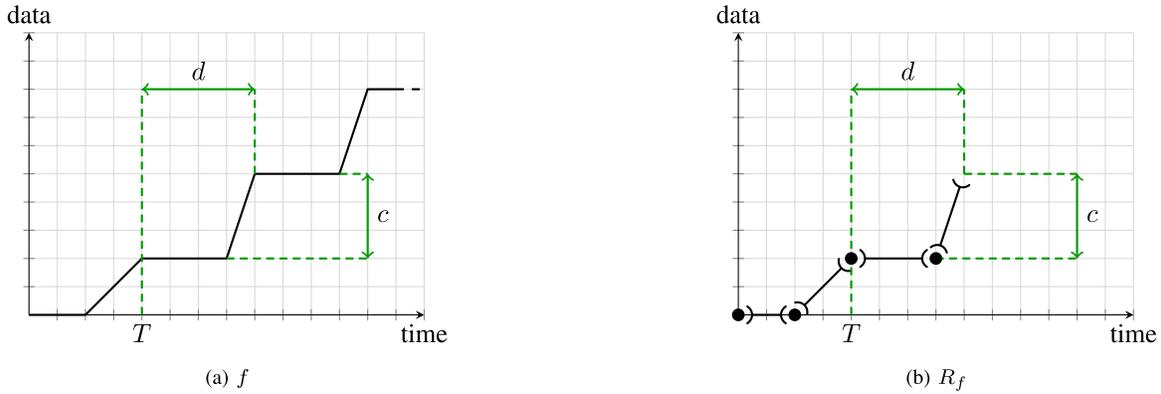

Fig. 1. Example of a continuous ultimately pseudo-periodic piecewise affine function $f$ and its representation $R_f$.

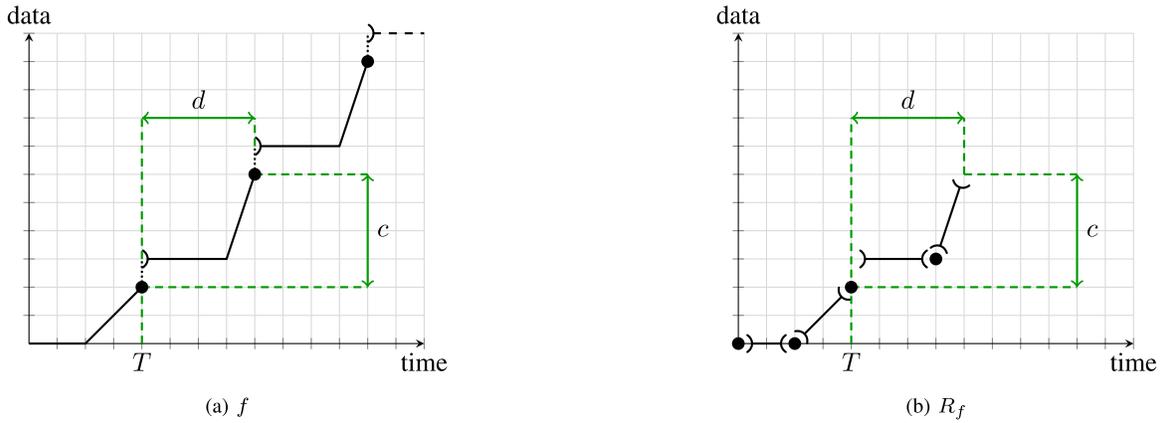

Fig. 2. Example of a left-continuous ultimately pseudo-periodic piecewise affine function $f$ and its representation $R_f$.

where we used the following shorthand notation for one-sided limits:

$$f\left(t_i^+\right) = \lim_{t \to t_i^+} f(t), \quad f\left(t_i^-\right) = \lim_{t \to t_i^-} f(t).$$

We use both points and open segments in order to easily model discontinuities. We will use the umbrella term *elements* to encompass both, when convenient.

*Definition 4 (Sequence):* Let a sequence $S_f^D$ be defined as on ordered set of elements $e_1, \ldots, e_n$ that alternate between points and segments and describe $f$ in finite interval $D$. Moreover, we define its *cardinality* $N\left(S_f^D\right)$ as the number of elements it contains.

For a function in $\mathcal{U}$, it is enough to store a representation of the initial transient part, i.e., interval $\mathcal{T} = [0, T[$, and of one period, i.e., interval $\mathcal{P} = [T, T + d[$. This entails storing a sequence describing the function in interval $[0, T + d[$. This is a finite amount of information. Figures 1 to 3 show examples of such functions. Accordingly, we call a *representation* $R_f$ of a function $f$ the tuple $(S, T, d, c)$, where $T, d, c$ are the values described above, and $S$ is the sequence defined in interval $[0, T + d[$.[2]

For example, for a rate-latency curve $\beta_{R,\theta}$ we have $T = \theta$, $d$ and $c$ can be arbitrary positive numbers such that $c/d = R$, and

[2]Whenever a sequence is defined over interval $[0, T+d[$, we omit to indicate the interval as a superscript for ease of notation.

$S$ is a list of four elements: point $(0, 0)$, segment $(0, \theta, 0, 0)$, point $(\theta, 0)$, segment $(\theta, \theta + d, 0, c)$.

Note that, given $R_f$, one can compute $f(t)$ for any $t \geq 0$, and also $S_f^D$ for any interval $D$. Furthermore, being finite, $R_f$ can be used as the data structure to represent $f$ in code. We remark that computing $S_f^D$, i.e., a sequence of $f$ over an arbitrary interval $D$, is a required step in most algorithms, for instance, to compare in the same interval two functions having different transients and/or periods.

As outlined in the Introduction, the aim of this paper is to present new techniques to reduce the computation times of NC computations. We therefore need to introduce the basic algorithms for NC operations, i.e., minimum, convolution and SAC, as well as determining the equivalence of two representations. A complete description of the algorithms for the above operations would be cumbersome, and would distract the reader from the main focus of this work. For this reason, we sketch here the basic results required for the understanding of the rest of the paper, and refer the interested reader to Appendix A for more details.

Given two representations $R_f$ and $R_g$ we can establish if $f = g$ through a linear comparison (element-by-element pairwise comparison) of sequences $S_f^D$, $S_g^D$, with $D = [0, \max(T_f, T_g) + \text{lcm}(d_f, d_g)[$. The alert reader can check that, if $f(t) = g(t) \; \forall t \in D$, then $f(t) = g(t) \; \forall t \geq 0$. The complexity of such comparison is then $\mathcal{O}\left(N\left(S_f^D\right) + N\left(S_g^D\right)\right)$.

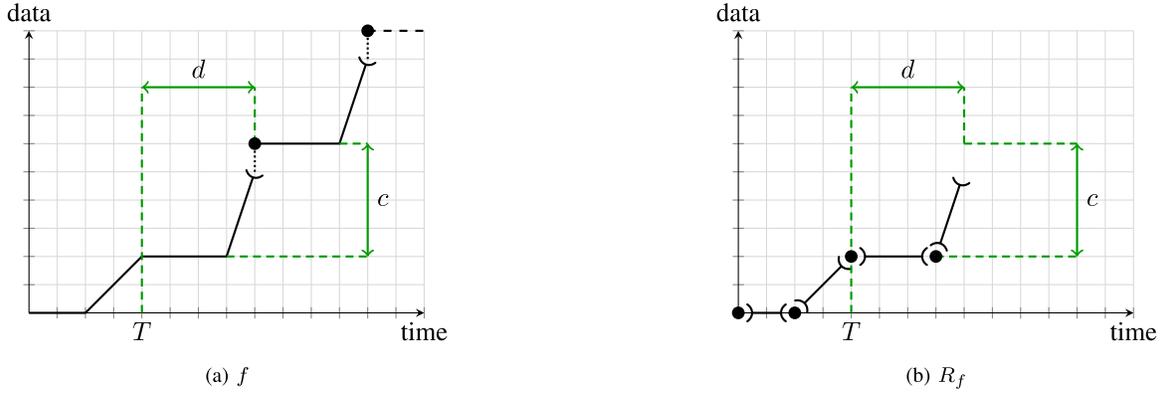

Fig. 3. Example of a right-continuous ultimately pseudo-periodic piecewise affine function $f$ and its representation $R_f$.

Binary NC operations (such as minimum and convolution) take as input the representations of operands and produce as an output the representation of the result. Given two functions $f$ and $g$ and a *generic* operator $*$, in order to compute $f * g$ we need an algorithm that computes $R_{f*g}$ from $R_f, R_g$, i.e., $(R_f, R_g) \to R_{f*g}$. This is generally done via the following steps:

1) compute valid parameters $T_{f*g}, d_{f*g}$ and $c_{f*g}$;
2) compute intervals $D_f$ and $D_g$, and the sequences $S_f^{D_f}$ and $S_g^{D_g}$, for the next step;
3) compute $(S_f^{D_f}, S_g^{D_g}) \to S_{f*g}$, i.e., provide an algorithm that computes the resulting sequence from the sequences of the operands (which we call *by-sequence* implementation of operator $*$);
4) return $R_{f*g} = (S_{f*g}, T_{f*g}, d_{f*g}, c_{f*g})$.

For the *minimum* operator, i.e., $m = f \wedge g$, the by-sequence algorithm is a linear comparison of sequences $S_f^{D_f}, S_g^{D_g}$, hence its complexity is $\mathcal{O}\left(N\left(S_f^{D_f}\right) + N\left(S_g^{D_g}\right)\right)$. However, intervals $D_f$ and $D_g$ (whose exact computation is reported in Appendix A) depend on numerical properties of $f$ and $g$. For instance, when $f$ and $g$ have different slopes $\rho_f, \rho_g$, and intersect at $\bar{t} \gg T_f + d_f, T_g + d_g$, these intervals are much larger than $[0, T_f + d_f[$ and $[0, T_g + d_g[$, respectively. Accordingly, it may well be that $N\left(S_f^{D_f}\right) \gg N(S_f)$ and/or $N\left(S_g^{D_g}\right) \gg N(S_g)$. This means that the computations involved in an operation may vary considerably based on the numerical properties of the operands. This issue will be recalled time and again throughout this paper.

For what concerns the *convolution* operation, we observe that in the general case it is not possible to compute the parameters and intervals of steps 1 and 2 *a priori*. To circumvent this, [24] proposes to decompose each operand $f$ into its transient and periodic parts, $f_t$ and $f_p$, each assuming value $+\infty$ outside their support (the set of $t \in \mathbb{Q}_+$ such that $|f(t)| < +\infty$ [24, p. 7]), so that $f = f_t \wedge f_p$. Therefore, convolution $f \otimes g$ can be decomposed as:

$$f \otimes g = (f_t \wedge f_p) \otimes (g_t \wedge g_p) = f_t \otimes g_t \wedge f_t \otimes g_p \wedge f_p \otimes g_t \wedge f_p \otimes g_p. \quad (6)$$

Section II-A highlights three types of partial convolutions: transient part with transient part; transient part with periodic part; periodic part with periodic part. For these partial convolutions, parameters and intervals can instead be computed. After all the partial convolutions in Section II-A have been computed, the end result can be obtained by taking the minimum of all partial results. A notable exception, which will be useful for this work, is when $\rho_f = \rho_g = \rho$. In this case, in fact, parameters and intervals can be computed a priori as follows:

$$T' = T_f + T_g + d'; \; d' = \mathrm{lcm}(d_f, d_g); \; c' = \rho \cdot d';$$
$$D_f = [0, T_f + 2 \cdot d'[; \; D_g = [0, T_g + 2 \cdot d'[.$$

The by-sequence algorithm consists in computing the convolutions of all the elements in the two sequences $S_f^{D_f}, S_g^{D_g}$, i.e., all $e_f \otimes e_g$ where $e_f \in S_f^{D_f}, e_g \in S_g^{D_g}$, and taking the lower envelope of the result. The complexity of this operation is $\mathcal{O}\left(N\left(S_f^{D_f}\right) \cdot N\left(S_g^{D_g}\right) \cdot \log\left(N\left(S_f^{D_f}\right) \cdot N\left(S_g^{D_g}\right)\right)\right)$. This complexity is heavily affected by the cardinalities of the sequences, which in turn are tied to $\mathrm{lcm}(d_f, d_g)$. Once again, it is possible that $N\left(S_f^{D_f}\right) \gg N(S_f), N\left(S_g^{D_g}\right) \gg N(S_g)$, for instance when $d_f, d_g$ are coprime.

The algorithm for computing SAC is based on Property 2. Consider the elements of $S_f$ $e_i$, $i = 1 \ldots n$, defined in $D_i = \{t_i\}$ (if $e_i$ is a point in $t_i$), or in $D_i = ]t_i, t_{i+1}[$ (if $e_i$ is a segment). Then, we can write a decomposition of $f$ into transient and periodic elements, $f = e_1^t \wedge \cdots \wedge e_l^t \wedge e_{l+1}^p \cdots \wedge e_n^p$, where[3]

$$e_i^t(t) = \begin{cases} f(t) & \text{if } t \in D_i, \\ +\infty & \text{otherwise}, \end{cases}$$

$$e_i^p(t) = \begin{cases} f(t + k \cdot d) & \text{if } t \in D_i + k \cdot d, k \in \mathbb{N}_0, \\ +\infty & \text{otherwise}. \end{cases}$$

Then, we have

$$\overline{f} = \overline{e_1^t} \otimes \cdots \otimes \overline{e_l^t} \otimes \overline{e_{l+1}^p} \cdots \otimes \overline{e_n^p}. \quad (7)$$

Thus, the SAC of $f$ is decomposed into SACs of points, open segments, periodic points and periodic open segments, for which algorithms are known [24]. This SAC computation is $\mathcal{NP}$-hard: the complexity grows exponentially with the number of elements in the representation. We remark that the above algorithm makes extensive use of the convolution operation.

[3]Throughout the paper, we use $\mathbb{N}$ to denote the set $\{1, 2, \ldots\}$ and $\mathbb{N}_0$ to denote the set $\{0, 1, \ldots\}$.

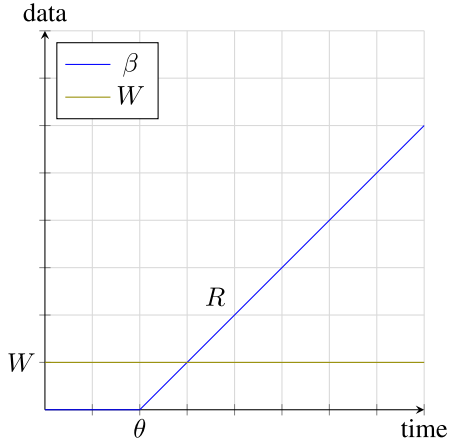
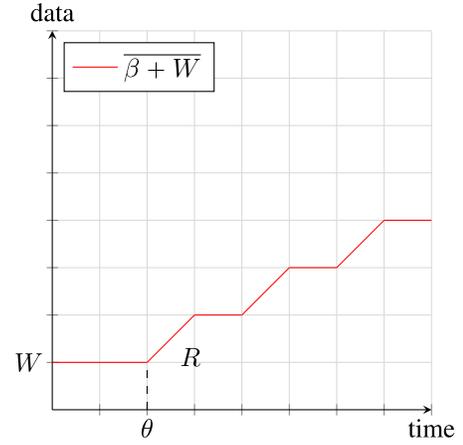

Fig. 4. SAC $\overline{\beta + W}$ (right) when $\beta$ is rate-latency and $W$ is the ordinate of a constant function (both shown on the left).

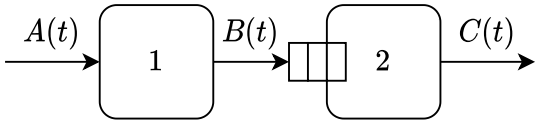

Fig. 5. Network with static window flow-control.

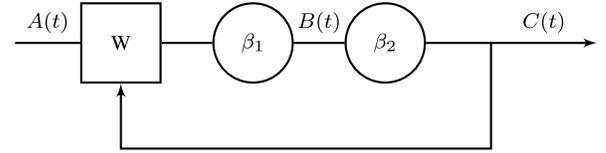

Fig. 6. NC model for network with static window flow-control. Circles represent service curve elements, whereas rectangles represent flow-control windows.

An exception is when $\beta$ is a rate-latency curve. In this case, given a *constant function*, i.e., a function $f$ such that $f(0) = 0$ and $\forall t > 0$, $f(t) = W$, SAC $\overline{\beta + f}$ can be computed in closed form [9]. Typically, $W$ is a buffer dimension. From now on, we will use shorthand $\overline{\beta + W}$, i.e., with the ordinate of the constant function rather than the function name itself, for better readability. As shown in Figure 4, the resulting function is not a rate-latency, but a staircase UPP function.

### B. Flow-Controlled Networks

NC can be used to model network elements having flow control [9, Chapter 4]. Consider the network in Figure 5, in which a flow traverses nodes 1 and 2, which have static flow-control due to the limited buffer in 2. Let $\beta_1$, $\beta_2$ be the service curves offered by nodes 1 and 2 to the flow that we are observing. Node 1 will thus serve that flow's traffic, with service curve $\beta_1$, only if there is already buffer space available in 2; in turn, the available part of this buffer space, whose size is $W$, depends on the ability of 2 to serve that flow's traffic with its own service curve $\beta_2$. We also assume that 1 is instantaneously aware of the current state of the buffer of 2. Thus, we can model the network as in Figure 6.

In order to compute an end-to-end service curve for a flow traversing the above system, we must first get rid of the feedback arc in the NC model, transforming it into a tandem. This is done by computing first the *equivalent service curve* of node 1, $\beta_1^{eq}$, such that

$$B(t) \geq (A \otimes \beta_1^{eq})(t).$$

The latter takes into account the reduction in the service brought on by the presence of the subsequent flow control. It is [9, Chapter 4]:

$$\beta_1^{eq} = \beta_1 \otimes \beta_{fc},$$
$$\beta_{fc} = \overline{\beta_1 \otimes \beta_2 + W}.$$

Then, the system offers to the flow an end-to-end service curve $\beta^{eq}$, so that

$$C(t) \geq (A \otimes \beta^{eq})(t).$$

$\beta^{eq}$ is equal to:

$$\begin{aligned}\beta^{eq} &= \beta_1^{eq} \otimes \beta_2 \\ &= \beta_1 \otimes \beta_2 \otimes \beta_{fc}.\end{aligned} \quad (8)$$

The extension of the above method to longer tandems is straightforward: consider for instance the tandem in Figure 7. Nodes 2 and 3 have limited buffers, of size $W_2$ and $W_3$. The resulting NC model is shown in the figure. To find the end-to-end service curve of the system, $\beta^{eq}$, we iterate the above methodology – starting from the rightmost node – and compute the following:

$$\begin{aligned}\beta_2^{eq} &= \beta_2 \otimes (\overline{\beta_2 \otimes \beta_3 + W_3}), \\ \beta_1^{eq} &= \beta_1 \otimes (\overline{\beta_1 \otimes \beta_2^{eq} + W_2}), \\ \beta^{eq} &= \beta_1^{eq} \otimes \beta_2^{eq} \otimes \beta_3.\end{aligned}$$

The above method is also illustrated in Figure 8. By expanding the expression of $\beta_1^{eq}$ we obtain the following:

$$\beta_1^{eq} = \beta_1 \otimes (\overline{\beta_1 \otimes \beta_2 \otimes (\overline{\beta_2 \otimes \beta_3 + W_3}) + W_2}). \quad (9)$$

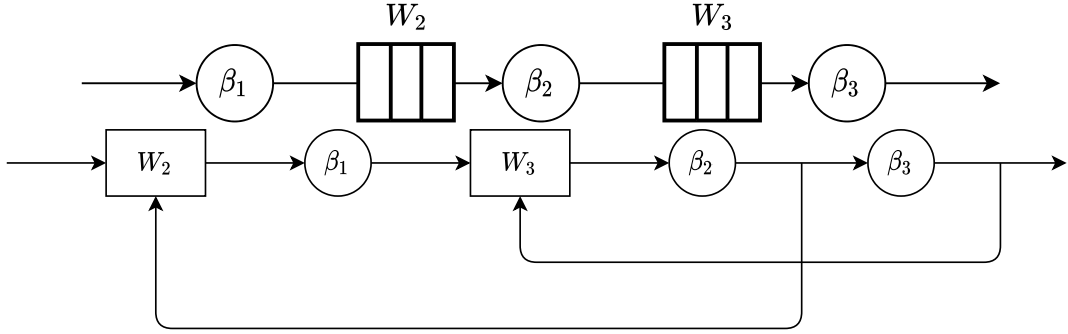

Fig. 7. A tandem of three flow-controlled nodes and its NC model.

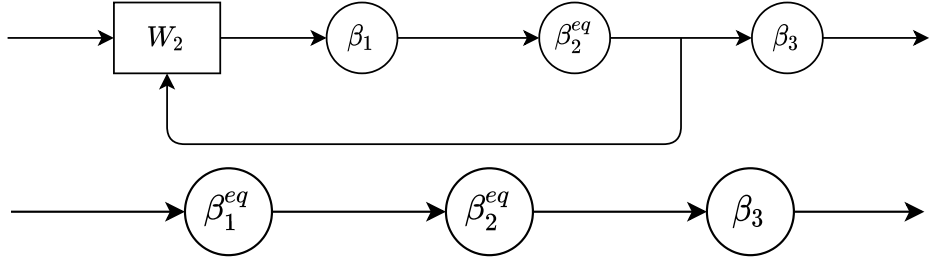

Fig. 8. Transformation of the NC model to an equivalent tandem.

Equation (9) includes a nested SAC. Even assuming the simplest case, i.e., that service curves $\beta_i$ are all rate-latency curves, the innermost SAC yields a staircase UPP function. The outer SAC must then be computed on a curve of this type, which is $\mathcal{NP}$-hard.

This method can be generalized to a tandem of $n$ nodes, as:

$$\beta_{n-1}^{eq} = \beta_{n-1} \otimes \overline{\beta_{n-1} \otimes \beta_n + W_n},$$
$$\beta_i^{eq} = \beta_i \otimes \overline{\beta_i \otimes \beta_{i+1}^{eq} + W_{i+1}},$$
$$\beta^{eq} = \left(\bigotimes_{i=1\ldots n-1} \beta_i^{eq}\right) \otimes \beta_n. \quad (10)$$

This method of analysis (henceforth: the *exact method*) therefore requires $\mathcal{O}(n)$ nested SACs for a tandem of $n$ nodes. All (except possibly one) are SACs of UPP curves which are non-trivial to compute (see (7)). Therefore, despite the apparent conciseness of (10), computing $\beta^{eq}$ via this method is computationally infeasible.

In [26], a property was proved that lower bounds $\beta_i^{eq}$ with a *convolution of SACs*:

$$\beta_i^{eq'} = \beta_i \bigotimes_{j=i}^{n-1} (\overline{\beta_j \otimes \beta_{j+1} + W_{j+1}}). \quad (11)$$

Then, an end-to-end service curve can be computed as:

$$\beta^{eq'} = \bigotimes_{i=1}^{n-1} \beta_i^{eq'} \otimes \beta_n$$
$$= \beta_1^{eq'} \otimes \beta_2^{eq'} \otimes \cdots \otimes \beta_{n-1}^{eq'} \otimes \beta_n$$
$$= \left(\beta_1 \otimes \bigotimes_{i=1\ldots n-1} \overline{\beta_i \otimes \beta_{i+1} + W_{i+1}}\right)$$
$$\otimes \left(\beta_2 \otimes \bigotimes_{i=2\ldots n-1} \overline{\beta_i \otimes \beta_{i+1} + W_{i+1}}\right) \otimes \cdots \otimes \beta_n$$
$$= \left(\bigotimes_{i=1\ldots n} \beta_i\right) \otimes \bigotimes_{i=1\ldots n-1} \overline{\beta_i \otimes \beta_{i+1} + W_{i+1}}. \quad (12)$$

The above is a consequence of each $\overline{\beta_i \otimes \beta_{i+1} + W_{i+1}}$ being sub-additive, thus $f \otimes f = f$.

Authors of [26] prove that:

$$\forall i = 1\ldots n-1, \beta_i^{eq} \geq \beta_i^{eq'}. \quad (13)$$

From the above, since convolution is isotonic, it follows that:

$$\beta^{eq} \geq \beta^{eq'}. \quad (14)$$

Computing $\beta^{eq'}$ via (12) (henceforth: the *approximate method*) is computationally more tractable – if all the SCs $\beta_i$ are rate-latency – because it does away with nested SACs. However, it still requires one to compute $\mathcal{O}(n)$ convolutions of UPP curves.

An exact expression for the service curve of the *first* node in a tandem of flow-controlled nodes has been derived in [29]. Its computation requires a chain of convolutions of sub-additive UPP curves, i.e., the same type whose computation we optimize in this paper.

## III. SYSTEM MODEL AND PROBLEM FORMULATION

Consider a flow traversing a tandem network of $n$ flow-controlled nodes. Each node $i$ offers to that flow a service curve $\beta_i$. After node $i$, $i < n$, there is a flow control window $W_{i+1}$. We initially assume that the flow-control is instantaneous, and implemented as described in Section II-B. We will come back to this issue at the end of this section. Our goal is to compute an end-to-end service curve of the above tandem network. This will allow one to compute a bound on the end-to-end delay and backlog, if the flow itself has an arrival curve. We want to be able to do this efficiently.

TABLE I
PARAMETERS OF THE EXAMPLE TANDEM NETWORK

|   | Exact | | | Approximate | | |
|---|---|---|---|---|---|---|
| $i$ | $\beta_i$ rate | $\beta_i$ latency | $W_{i+1}$ | $\beta_i$ rate | $\beta_i$ latency | $W_{i+1}$ |
| 1 | 8  | 5 | 3 | 21 | 15 | 23 |
| 2 | 11 | 7 | 7 | 30 | 17 | 29 |
| 3 | 12 | 4 | 3 | 7  | 27 | 20 |
| 4 | 1  | 5 |   | 21 | 20 |    |

TABLE II
COMPUTATIONAL RESULTS, EXACT METHOD

| comp. time of $\overline{\beta_2 \otimes \beta_3^{eq} + W_3}$ | 6 h 24 m |
|---|---|
| $N\left(\beta_2 \otimes \beta_3^{eq} + W_3\right) \to N\left(\overline{\beta_2 \otimes \beta_3^{eq} + W_3}\right)$ | $10 \to 10600$ |
| comp. time of $\overline{\beta_1 \otimes \beta_2^{eq} + W_2}$ | $> 24$ h |
| $N\left(\beta_1 \otimes \beta_2^{eq} + W_2\right) \to N\left(\overline{\beta_1 \otimes \beta_2^{eq} + W_2}\right)$ | $unknown$ |

TABLE III
COMPUTATIONAL RESULTS, APPROXIMATE METHOD

| comp. time of $\beta_{\{1,3\}} = \overline{\beta_1 \otimes \beta_2 + W_2} \otimes \overline{\beta_2 \otimes \beta_3 + W_3}$ | 0.14 s |
|---|---|
| $N\left(\overline{\beta_1 \otimes \beta_2 + W_2}\right), N\left(\overline{\beta_2 \otimes \beta_3 + W_3}\right) \to N(\beta_{\{1,3\}})$ | $6, 6 \to 270$ |
| comp. time of $\beta_{\{1,4\}} = \beta_{\{1,3\}} \otimes \overline{\beta_3 \otimes \beta_4 + W_4}$ | 6 h 13 m |
| $N(\beta_{\{1,3\}}), N\left(\overline{\beta_3 \otimes \beta_4 + W_4}\right) \to N(\beta_{\{1,4\}})$ | $270, 6 \to 1456$ |

We first show that computing an end-to-end service curve, whether the exact or the approximate one, incurs state explosion and may require very long computation times, even when $n$ is small (e.g, three nodes).

We do this using a four-hop tandem of flow-controlled nodes as an example. We assume that nodes have rate-latency service curves, $\beta_i = \beta_{R_i,\theta_i}$, and their parameters are those in Table I. We need to use two different sets of parameter values to better highlight the issues – and, later, the impact of the optimizations. In fact, the length and/or feasibility of the computations do depend on the parameter values. We found that a setting that can be solved with the exact method is often trivial with the approximate one, and a setting that is hard with the approximate method is often computationally intractable with the exact one.

We attempt to compute the end-to-end equivalent service curves, via the exact and approximate methods, using the algorithms described in [24], on a desktop PC equipped with an Intel Core i9-9900, 16 GB of DRAM @3200 MHz, Windows 10 (the above system, included the software we run, is described in more detail in Section V).

We report computational results in, respectively, Table II and Table III, where we highlight both the representation size of the results and the time it takes to compute them. The alert reader will notice that the results reported in the tables are intermediate results towards the equivalent end-to-end service curves via (10) and (12), respectively. We cap computation times at 24 hours.

The above results show that computation times are non-trivial, and that state explosion does occur, even with the approximate method. As already outlined, we cannot abate these computation times by working on *compact domains*, as suggested in [28]. In fact, that method requires that the service curves involved are *super-additive*. In our model, the operands of these tough convolutions are instead *sub-additive*, because they are the result of SACs. We are not aware of any method that allows one to limit the domains in this case.

A simple, but crude approach to compute an *approximated* end-to-end service curve would be to *lower bound* each resulting UPP curve with a rate-latency curve. This approximation would have to be used after each SAC in (10), or for each term in (12). This would certainly make computations considerably faster, but may entail a considerable loss of accuracy.

We exemplify this using a simple UPP curve $\beta = \beta_{R,\theta} \otimes \overline{\beta_{R,\theta} + h}$, with $R \cdot \theta > h$. In this case, such lower bound $\lfloor \beta \rfloor_{rl}$ would have $\theta_{lb} = \theta$, $R_{lb} = \frac{h}{\theta}$, and the error introduced by it is upper bounded by $\theta - \frac{h}{R}$. As shown in Figure 9, the impact of such error on the end-to-end delay depends on the characteristic of the input traffic. Notably, small messages would incur relatively larger penalty than large messages, and the loss in accuracy would be non-negligible.

Our approach to gaining efficiency is to abate both the number and the computation time *of the convolutions of UPP curves*. This operation, in fact, lies at the core of both the exact and the approximate methods (recall that the SAC of a UPP curve can be computed as a convolution of elements, as explained in Section II-A). Reducing their number and making them as fast as possible is therefore going to make both methods more efficient. We do this *without introducing approximations*: our computations are always exact. In the next section, we show how we accomplish this, leveraging both representation minimization and algebraic tools: first, we show that *minimizing the representation $R_f$ of the functions $f$* involved in the operations may provide remarkable benefits. Then, we present three theorems that can be used to reduce the computation time of convolutions, leveraging sub-additivity of the operands.

Before introducing our contribution, we spend a few words discussing the generality of our model. With flow-controlled networks, different models can be envisaged as far as:

1) the exact place where flow-control stops traffic when the flow control window is closed, w.r.t. to the service curve modelling the sending node. For example, this may be an input buffer of the sending node, or an output buffer instead. The alert reader can check that using one or the other will lead to slightly different expressions for both the exact and the approximate end-to-end service curves. However, they will still be of the same type as (10) and (12), respectively, i.e., with either nested SACs or convolutions of sub-additive UPPs. Therefore, any computation issue that we address throughout this paper will still be present;
2) whether or not the return path, i.e., the flow-control credit feedback, is instantaneous. Depending on how such feedback is implemented, other models may, for instance,

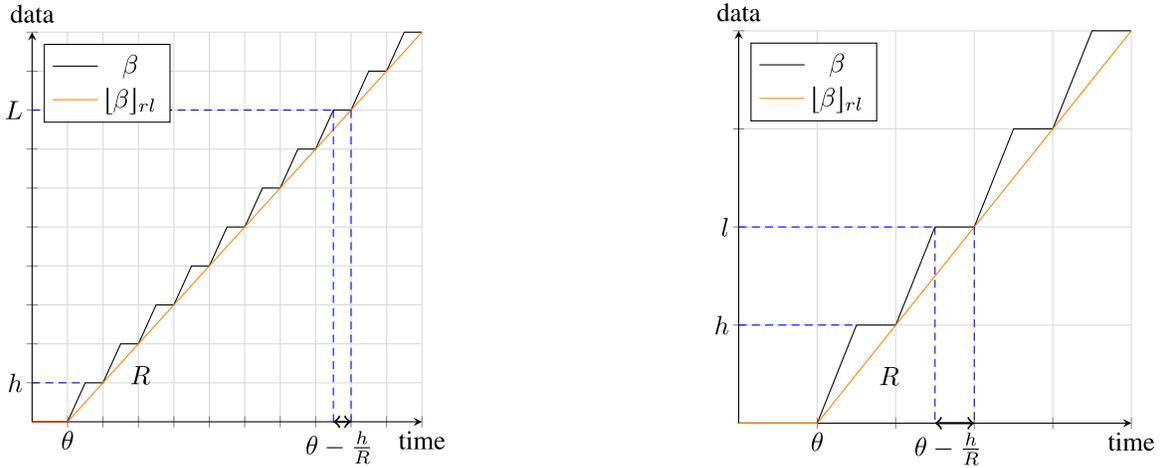

(a) Delay bound for a long message of length $L$. The one obtained using $\lfloor \beta \rfloor_{rl}$ is overestimated by $\theta - \frac{h}{R}$, 6% w.r.t. the one obtained using $\beta$.

(b) Delay bound for a short message of length $l$. The one obtained using $\lfloor \beta \rfloor_{rl}$ is overestimated by $\theta - \frac{h}{R}$, 20% w.r.t. the one obtained using $\beta$.

Fig. 9. Delay overestimation introduced by lower-bounding a staircase UPP curve with a rate-latency curve.

include a service curve on the return path as well. Again, this does not change the structure of the expressions that we seek to compute efficiently.

Thus, only the model of Figure 6 will be considered henceforth for simplicity.

## IV. CONTRIBUTION

We present our two contributions separately. First, we show how to minimize the representation of a function, using an inexpensive algorithm. Then, we show theorems that reduce the cost of convolutions.

### A. Representation Minimization

As discussed, given a representation $R_f = (S, T, d, c)$ of function $f$, its cardinality $N(S)$ and parameters $d$ and $T$ are the main factors for the algorithmic complexity of operations involving it. A first way to abate computation times is therefore to find the *minimal* representation of $f$.

We say that two representations $R_f$ and $R_g$ are *equivalent* if they represent the same function, i.e., $\forall t \geq 0, f(t) = g(t)$. A *minimal* representation $\tilde{R}$ is such that, given any equivalent representation $R$, then $N(\tilde{S}) \leq N(S)$.

Unfortunately, the generic algorithms for operations on UPP curves (described in [24], [25] and recalled in Section II-A and Appendix A) do not yield minimal representations, even when the representations of their operands are minimal. The steps described in Section II-A, in fact, compute the smallest values that can be formally proved to be valid *a priori*, with no knowledge of the shape of the result. These values can be much larger than those of a minimal representation.

A simple example is given in Figure 10. Starting from the parameters of the operands $f$ and $g$, the algorithm computes $T = 7$ for the result $f \wedge g$. However, we can see that the result is actually a rate-latency curve that can be described with $T = 5$. This phenomenon – that we have just exemplified using a minimum operation – affects convolution as well, and it is especially impactful when many convolutions are required, such as in a SAC or in (11), where the result of one is in fact the operand of the next. In fact, we recall that the cost of the convolution is superquadratic with the size of the extended representations of the operands (Section II-A).

Note that there is no efficient way – that we know of – to predict the minimum representation *a priori*, i.e., before the operation is computed. This is basically because the result depends on unpredictable numerical properties of the operands (e.g., the segment endpoints). We therefore introduce an algorithm to minimize the representation *a posteriori*, i.e., after the result of the operation has been computed. We will show later that minimization is computationally cheap, and may yield considerable speedups.

We say that $t_b$ is a *breakpoint* of $f$ if $f$ is non-differentiable in $t_b$, i.e., either of the following is true:

- $f$ has a discontinuity at $t_b$;
- the rates of $f$ in $t_b^-$ and $t_b^+$ are different.

A first thing to do is to ensure that the sequence in a representation is *well-formed*. We say that $S$ is a *well-formed* sequence if the abscissa of any point in $S$ is a breakpoint of $f$. In other words, in a well-formed sequence there are no *unnecessary* points.[4]

As we anticipated, the generic algorithms for minimum and convolution may not yield well-formed sequences, even when the sequences of their operands are well-formed. However, recovering well-formedness only takes a simple $\mathcal{O}(N(S))$ check of the resulting sequence $S$, to find segment-point-segment triplet that can be merged, i.e., replaced with a single segment. From now on, we will therefore assume that sequences are well-formed, without loss of generality.

We describe below a minimization algorithm consisting of two phases:

---

[4]An exception must be made at $T$, where a point has to be inserted regardless, marking the end of the transient and the beginning of the periodic part, because this simplifies the implementation. Such an exception has no impact on the rest of our discussion.

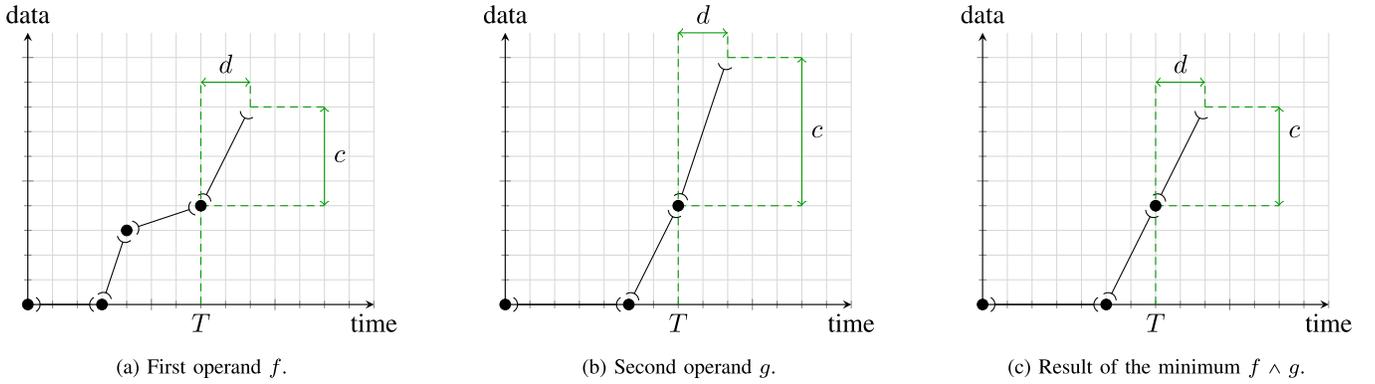

Fig. 10. Example of non-minimal result of a minimum operation.

- minimization of the period;
- minimization of the transient.

Hereafter, we denote with $S^\mathcal{T}$ the *transient part* of a sequence $S$ (i.e., in interval $\mathcal{T} = [0, T[$) and with $S^\mathcal{P}$ its *periodic part* (i.e., in interval $\mathcal{P} = [T, T+d[$).

*1) Minimization of the Period:* We set to finding the minimal period $\tilde{d}$, defined as follows:

*Definition 5:* A minimal period $\tilde{d}$ for $f$ is such that $R_f = (S, T, \tilde{d}, \tilde{c})$ is a representation of $f$, and there exists no $q \in (0, 1) \subset \mathbb{Q}$ such that

$$f(t + q \cdot \tilde{d}) = f(t) + q \cdot \tilde{c} \qquad \text{for all } t \geq T.$$

Period minimization is only relevant if the function is not *Ultimately Affine* (UA). Graphically, a UA function ends with a half-line. More formally:

*Definition 6:* A function $f \in \mathcal{U}$ is UA if

$$\forall t \geq T, \ \forall \delta \geq 0, \qquad f(t + \delta) = f(t) + \rho_f \cdot \delta.$$

The above definition is equivalent to saying that $f$ has no breakpoint for any $t > T$. Conversely, any $f \in \mathcal{U}$ having a breakpoint $t_b > T$ is not UA.

Note that it is important that inequality $t > T$ is strict. In fact, whether $T$ itself is a breakpoint or not may depend on the transient behavior. On the other hand, we are interested in the periodic behavior. Therefore, we check if $T + d$ is a breakpoint, i.e., if point $(T, f(T))$, repeated after a period in $(T + d, f(T) + c)$, breaks the linear behavior between one pseudo-period and the next. Figure 11 shows two examples to illustrate the above. For this reason, in the following we will focus on the breakpoints in $]T, T + d]$.

A UA function has no minimal period. In fact, its period has an arbitrary length $d > 0$. Accordingly, its $S^\mathcal{P}$ only consists of point $(T, f(T))$ and a segment of length $d$ and slope $\rho_f$, hence $c = \rho_f \cdot d$. For UA functions, then, there is just nothing to do. Conversely, any $f$ which is not UA has a minimal period. In fact, call $t_b$ the *leftmost* breakpoint such that $t_b > T$ (we know that there is at least one): then, the interval $]T, T + d]$ must include $t_b$, since it must include at least one breakpoint if $f$ is not UA. Then, $d \geq t_b - T > 0$.

The next question, then, is what characterizes non-minimal periods and how we can find the minimal one, given a representation. It is fairly obvious that non-minimal periods are integer multiples of the minimal one (this can also be proved formally, see Appendix B). Given $R_f = (S, T, d, c)$, if $f$ also admits a smaller period $d'$, it must hold that $d/d' = c/c' = p$, where $p \in \mathbb{N}$. Such $p$ also divides $S^\mathcal{P}$ in $p$ *matching parts*, i.e., such that $\forall t \in [T, T + \frac{d}{p}[$ and $k \in \mathbb{N}$ it holds that $f(t + k \cdot \frac{d}{p}) = f(t) + k \cdot \frac{c}{p}$. Hence, we call $p$ a *divisor of* $S^\mathcal{P}$. We exemplify this in Figure 12. Figure 12a shows $f$ and its representation, with breakpoints in $]T, T+d]$ highlighted as circles. Figure 12b shows that $d/3$ is also a (minimal) period, and that – accordingly – $S^\mathcal{P}$ consists of $p = 3$ consecutive replicas of a smaller periodic part $S^{\mathcal{P}'}$, which is highlighted in red in the figure. Thus, in order to minimize the period of a non-UA function, we need to find the possible divisors of $S^\mathcal{P}$, i.e., to test if the latter is in fact the juxtaposition of matching parts. This can be done efficiently by observing the following.

*Lemma 1:* Let $f \in \mathcal{U}$ be non UA, and let $b$ be the number of its breakpoints in $]T, T + d]$. Then, if $p \in \mathbb{N}$, $p > 1$, is a divisor of $S^\mathcal{P}$, it is also a divisor of $b$.

*Proof:* Let $\tilde{d}$ be the minimal period for $f$, and let $\tilde{b}$ be the number of breakpoints in $]T, T + \tilde{d}]$. Now, by definition, $d = p \cdot \tilde{d}$ for some $p \in \mathbb{N}$. By construction, then, if $p > 1$, $S^\mathcal{P}$ consists of $p$ matching parts, hence $p$ is a divisor of $S^\mathcal{P}$. By Equation (5), if $t_b \in ]T, T + \tilde{d}]$ is a breakpoint, then $t_b + \tilde{d}$ is also a breakpoint. Then, it is $b = p \cdot \tilde{b}$, i.e., $p$ is also a divisor of $b$. □

For instance, in Figure 12b, $p = 3$ is a divisor of $S^\mathcal{P}$, and there are in fact 6 breakpoints in $]T, T + d]$.

Lemma 1 states that, in order to minimize the period, we can limit ourselves to testing if *the divisors of $b$* are also divisors of $S^\mathcal{P}$.

This allows us to formulate an efficient algorithm that minimizes the period $d$ of a representation $R_f$:

- Count $b$ as the number of breakpoints of $f$ in $]T, T + d]$;
- Find the prime factorization of $b$;
- Exhaustively test if the prime factors of $b$ are also divisors of $S^\mathcal{P}$: if they are, update $S^\mathcal{P}$, $d$ and $c$ accordingly.

We exemplify this algorithm on the function in Figure 12. From Figure 12a we observe that $b = 6$, whose prime factorization is $2 \cdot 3$. Therefore, we test these two primes as possible divisors of $S^\mathcal{P}$, as shown in Figure 13. Testing a factor $p$ consists, in general, in dividing the sequence in $p$ parts, defined in $[T + i \cdot \frac{d}{p}, T + (i+1) \cdot \frac{d}{p}[$ for $i = 0 \ldots p - 1$,

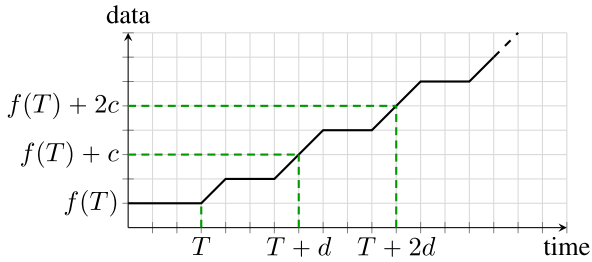

(a) Example of function $f$ with a breakpoint in $T$, but not in $T + k \cdot d$.

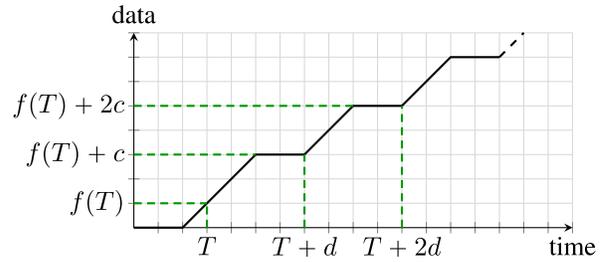

(b) Example of function without a breakpoint in $T$, but with breakpoints in $T + k \cdot d$.

Fig. 11. Breakpoints in $T$ vs. $T + d$.

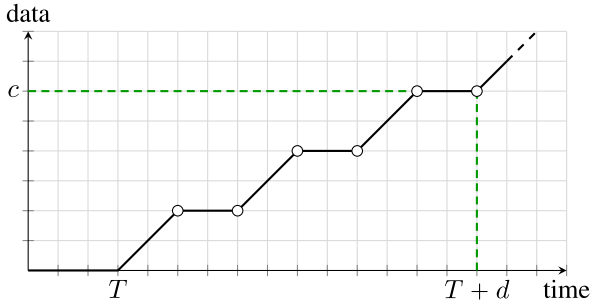

(a) $f$ with breakpoints in $]T, T + d]$ highlighted as circles.

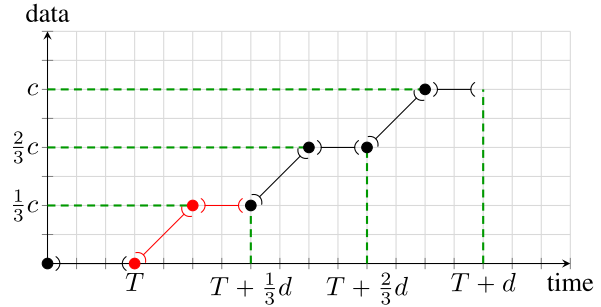

(b) Factorization of $S^\mathcal{P}$ with $p = 3$.

Fig. 12. Example of factorization of the pseudo-periodic part. We can replace $d$, $c$ and $S^\mathcal{P}$, defined in $[T, T + d[$, with, respectively, $\frac{d}{3}$, $\frac{c}{3}$, and $S^{\mathcal{P}'}$, defined in $[T, T + \frac{d}{3}[$ and highlighted in red. The latter is an equivalent, but more compact, representation of $f$.

and checking whether, after shifting them down by $\frac{d}{p}$ and left by $\frac{c}{p}$, they all match. Figure 13a shows that the test with $p = 2$ fails, whereas Figure 13b shows that the test with $p = 3$ succeeds. After a division succeeds, it is convenient to immediately replace $S^\mathcal{P}$, $d$ and $c$ with their smaller equivalents $S^{\mathcal{P}'}$, $\frac{d}{p}$ and $\frac{c}{p}$, so that the upcoming tests with other factors of $b$ are more efficient. In particular, $N(S^{\mathcal{P}'}) < N(S^\mathcal{P})$. The test is run exhaustively for all prime factors of $b$. If a prime factor $p$ has multiplicity $m > 1$, it is tested as a divisor of $S^\mathcal{P}$ up to $m$ times.

Obtaining number $b$ requires counting breakpoints of $f$ in $]T, T + d]$, which is a simple $\mathcal{O}\left(N(S^\mathcal{P})\right)$ check. To find the prime factorization of $b$, we will need the prime numbers in $2 \ldots \sqrt{b}$. Computing primes until a given $x$ is something that can be done offline – we use an offline list of 1000 primes in our implementation, which is enough for periods exceeding 62 millions. Lastly, testing if a prime factor $p$ is a divisor of $S^\mathcal{P}$ entails a linear comparison between its parts. Let $n_p$ be the number of prime divisors of $b$. In the worst-case, the algorithm will test, unsuccessfully, all $n_p$ divisors, thus the complexity of this last step is $\mathcal{O}\left(n_p \cdot N(S^\mathcal{P})\right)$.

*2) Minimization of the Transient:* In a non-minimal representation, the period start $T$ can be overestimated, making the transient part longer than strictly necessary. This algorithm aims at removing this excess transient by bringing forward the period start. We exemplify this process starting from the representation in Figure 14.

As a first step, which we call *by-period*, we check if the rightmost end of the transient part contains sequences that match with the (already minimized) periodic part itself – and remove them, in case. In the example of Figure 14 we can see that the representation is equivalent to the one in Figure 15.

We can obtain this result algorithmically by comparing the sequence in $[T, T + \tilde{d}[$ with the transient sequence immediately before, i.e., in $[T - \tilde{d}, T[$. If the two are matching, then the period start can be brought forward to $T' = T - \tilde{d}$, while the other parameters stay the same. This operation removes a period's worth of elements from $S^\mathcal{T}$. We repeat this process iteratively until the comparison fails. The end result is a reduction of the representation by a number of periods $k \in \mathbb{N}_0$, and an earlier period start $T' = T - k \cdot \tilde{d}$.

As a second step, which we call *by-segment*, we test if *parts* of a period, instead of whole periods, can be found at the right end of the transient part. In the example we can see that the representation of Figure 15 is equivalent to the one in Figure 16.

We can algorithmically obtain this result by comparing the last pair (point, segment) of the periodic part, say in $[T + \tilde{d} - l, T + \tilde{d}[$, thus of length $l$, with the transient part of the same length immediately before the period start, thus in $[T - l, T[$. If the two are matching, then the period start is brought forward to $T' = T - l$, while the other parameters stay the same.

While this appears close to the *by-period* step, an important difference is that $S^\mathcal{P}$ needs also be altered as a result (although $\tilde{d}$ will remain the same).

The above steps are repeated until no further changes can be made. Transient minimization can also be applied to UA functions. For these, one should just check if the tail of the transient

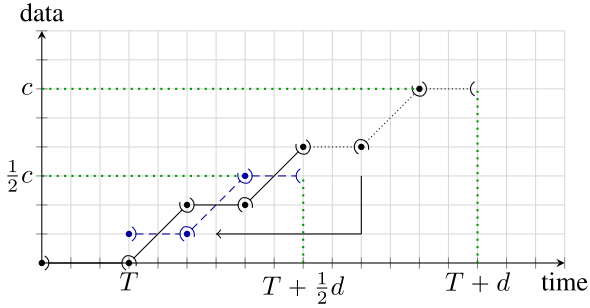
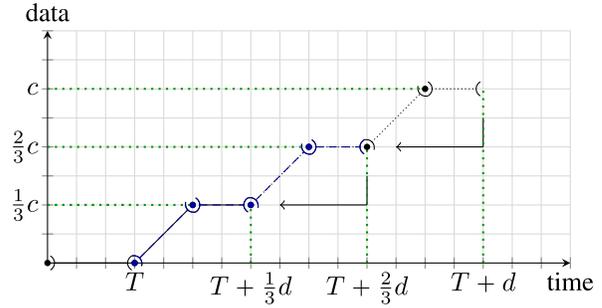

(a) Test of prime factor 2. The last half is translated (down by $\frac{1}{2}c$, left by $\frac{1}{2}d$) on top of the first half. Since they do not match, 2 is not a divisor of $S^{\mathcal{P}}$.

(b) Test of prime factor 3. The last third is translated (down by $\frac{1}{3}c$, left by $\frac{1}{3}d$) on top of the second one, and then again on top of the first one. As they all match, 3 is a divisor of $S^{\mathcal{P}}$.

Fig. 13. Example of the factorization algorithm.

TABLE IV
COMPUTATIONAL RESULTS, EXACT METHOD

|  | unoptimized results | with minimization |
|---|---|---|
| comp. time of $\overline{\beta_2 \otimes \beta_3^{eq} + W_3}$ | 6 h 24 m | 6 s |
| $N\left(\beta_2 \otimes \beta_3^{eq} + W_3\right) \to N\left(\overline{\beta_2 \otimes \beta_3^{eq} + W_3}\right)$ | $10 \to 10600$ | $10 \to 10$ |
| comp. time of $\overline{\beta_1 \otimes \beta_2^{eq} + W_2}$ | $> 24$ h | 13 s |
| $N\left(\beta_1 \otimes \beta_2^{eq} + W_2\right) \to N\left(\overline{\beta_1 \otimes \beta_2^{eq} + W_2}\right)$ | *unknown* | $14 \to 6$ |

TABLE V
COMPUTATIONAL RESULTS, APPROXIMATE METHOD

|  | unoptimized results | with minimization |
|---|---|---|
| comp. time of $\beta_{\{1,3\}} = \overline{\beta_1 \otimes \beta_2 + W_2} \otimes \overline{\beta_2 \otimes \beta_3 + W_3}$ | 0.14 s | 0.11 s |
| $N\left(\overline{\beta_1 \otimes \beta_2 + W_2}\right), N\left(\overline{\beta_2 \otimes \beta_3 + W_3}\right) \to N(\beta_{\{1,3\}})$ | $6, 6 \to 270$ | $6, 6 \to 42$ |
| comp. time of $\beta_{\{1,4\}} = \beta_{\{1,3\}} \otimes \overline{\beta_3 \otimes \beta_4 + W_4}$ | 6 h 13 m | 13.47 s |
| $N\left(\beta_{\{1,3\}}\right), N\left(\overline{\beta_3 \otimes \beta_4 + W_4}\right) \to N(\beta_{\{1,4\}})$ | $270, 6 \to 1456$ | $42, 6 \to 6$ |

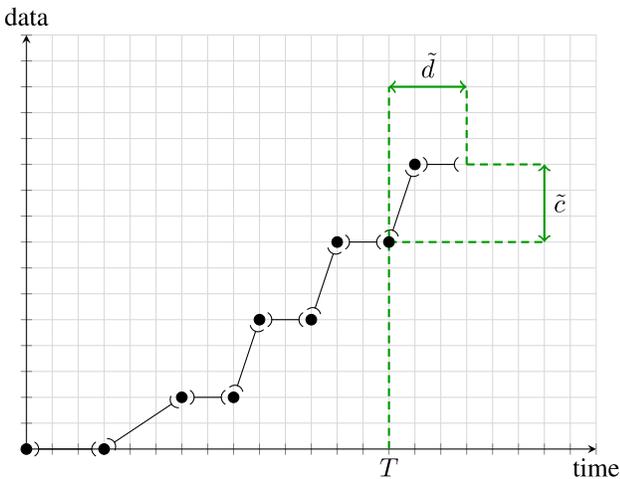

Fig. 14. Example of a non-minimal representation.

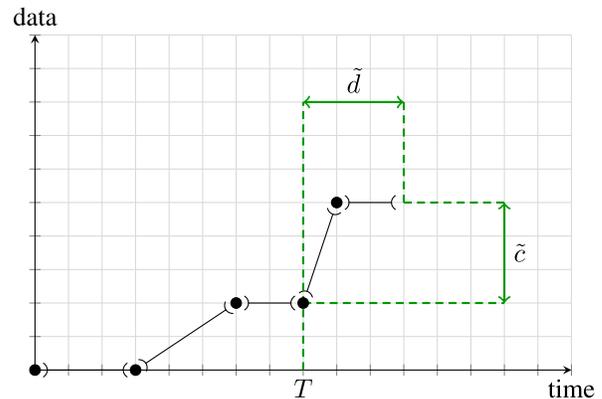

Fig. 15. Representation reduced by a whole number of periods.

is aligned with the period half-line. Our transient minimization algorithm always achieves the minimum $T$, if one exists. In fact, some functions do not admit a *minimum* $T$. Figure 17 shows a function which is not right-continuous in $T_L$, and whose periodic part must start *after* $T_L$ (recall that Equation (5) includes a *weak* inequality), hence admits no *minimum* $T$. In this particular case, our by-segment step yields the representation in Figure 17. We observe that using any $T'$ in $]T_L, T]$, by removing a fragment of the rightmost segment in the periodic part, would yield an equivalent representation with the same $N(S)$ anyway.

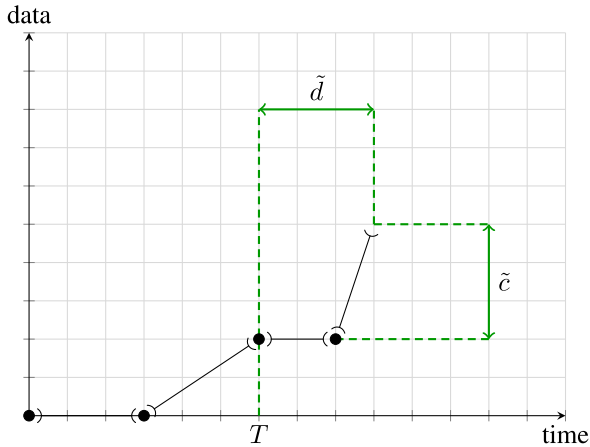

Fig. 16. Representation reduced by a segment, altering the pseudo-period sequence.

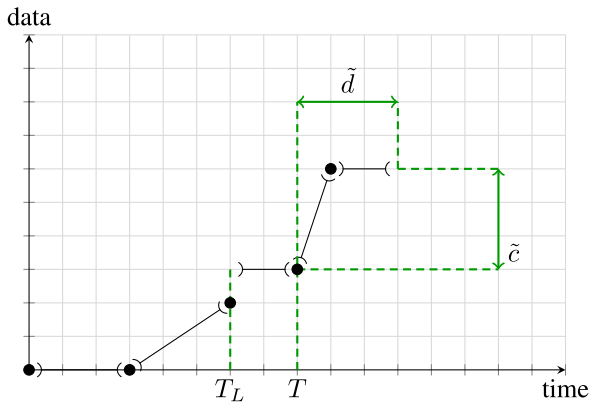

Fig. 17. Example of function with a right-discontinuity before the period start: $T_L$ is an infimum for $T$, but not its minimum.

As the linear comparisons involve, at most, the entire $S^T$, the cost of this algorithm is $\mathcal{O}\left(N\left(S^T\right)\right)$. Thus the cost of the entire representation minimization algorithm is $\mathcal{O}\left(N\left(S^T\right) + n_p \cdot N\left(S^\mathcal{P}\right)\right)$.

With reference to the example presented in Section III, we repeated the same computations, this time adding representation minimization in between each operation (both minima and convolutions).

The new results are in Table IV and Table V, which highlight both speedups and reductions in representation size up to three orders of magnitude.

An important aspect that links both is that a larger representation size translates directly to a higher memory occupancy during computations. As the occupied memory approaches the maximum allowed by the computer system, the performance is also affected. We do believe that the reason why some SACs do not terminate with the exact method is that they end up occupying all the available memory, hence disk swaps start kicking in.

### B. Efficient Convolutions of Sub-Additive Functions

In a convolution the resulting period grows like the lcm of the period of the operands, and the complexity of the by-sequence algorithm is superquadratic with the length of the sequences. Thus, it is possible to find instances where a *single* convolution may take very long.

It is however possible to leverage sub-additivity to reduce the complexity of the convolution. To the best of our knowledge, this has never been observed before. We recall that the convolutions we aim to optimize involve functions $\in \mathcal{U}$ which are sub-additive and such that $f(0) = 0$ by construction (they are in fact the result of SAC operations). In the following theorems and properties, however, we will reduce to a minimum the set of required hypotheses, in the interest of generality.

We first observe that *dominance* can be leveraged to abate the complexity of convolutions. We say that $g$ *dominates* $f$ if $\forall t \geq 0, g(t) \geq f(t)$. In this case:

*Theorem 1 (Convolution of Sub-Additive Functions With Dominance):* Let $f$ and $g$ be functions $\in \mathcal{U}$ such that $g(0) = 0$, $\forall t, g(t) \geq f(t)$, $f$ is sub-additive, and $f \otimes g$ is well-defined.[5] Then,

$$f \otimes g = f. \qquad (15)$$

*Proof:* Since $f$ is sub-additive, $\forall u, s, f(u) + f(s) \geq f(u+s)$. Then, for any $t = u+s$, $f(u) + g(s) \geq f(u) + f(s) \geq f(t)$. Thus, $(f \otimes g)(t) = \inf_{u+s=t}\{f(u) + g(s)\} = f(t) + g(0) = f(t)$. □

In order to apply this theorem algorithmically, we first need to compare $f$ and $g$. Dominance can be verified by checking statements $f = f \wedge g$ and $g = f \wedge g$, using the equivalence algorithm. Both the minimum and the equivalence check have linear costs as discussed in Section II-A. When either is true, Theorem 1 allows us to bypass the convolution altogether, which is instead superquadratic. Note that this theorem (as well as the following one) requires only the *dominated* function $f$ to be sub-additive, whereas the dominant function $g$ can have any shape, as long as $g(0) = 0$.

When dominance does not hold, we can test a weaker property, *asymptotic dominance*. We say that $g$ *dominates* $f$ *asymptotically* if $\exists\, t^* > 0$ such that $\forall t \geq t^*, g(t) \geq f(t)$. Note that $\rho_g > \rho_f$ is a sufficient condition for this to occur, but not a necessary one. In this case, we can resort to a "simpler" convolution as follows:

*Theorem 2 (Convolution of Sub-Additive Functions With Asymptotic Dominance):* Let $f$ and $g$ be functions $\in \mathcal{U}$ such that $f(0) = g(0) = 0$, $\forall t \geq t^*, g(t) \geq f(t)$, $f$ is sub-additive and $\forall t \geq 0, f(t) > -\infty$.

Let $g = g_a \wedge g_b$ be a decomposition of $g$ where:

$$g_a(t) = \begin{cases} g(t) & \text{if } t \in [0, t^*[ \\ +\infty & \text{if } t \geq t^* \end{cases}, \quad g_b(t) = \begin{cases} 0 & \text{if } t = 0 \\ +\infty & \text{if } t \in ]0, t^*[ \\ g(t) & \text{if } t \geq t^* \end{cases}.$$

Then

$$f \otimes g = f \otimes g_a \wedge f.$$

*Proof:* Decompose $g$ as per the hypothesis. Then:

$$f \otimes g = f \otimes (g_a \wedge g_b)$$
$$= f \otimes g_a \wedge f \otimes g_b.$$

---
[5]As mentioned in [24, p. 7], convolution $f \otimes g$, with $f, g \in \mathcal{U}$, is not defined if there exist $t_1, t_2$ such that $f(t_1) = +\infty$ and $g(t_2) = -\infty$, or vice versa (i.e., both are infinite, with opposite signs).

For the latter part, we observe that $\forall t, g_b(t) \geq f(t)$, and that $g_b(0) = 0$. We can therefore apply Theorem 1, for which $f \otimes g_b = f$. Thus:

$$f \otimes g = f \otimes g_a \wedge f \otimes g_b$$
$$\stackrel{(15)}{=} f \otimes g_a \wedge f.$$

□

If $g$ dominates $f$ only asymptotically, then $f$ is above $g$ at some point, but will eventually fall below it. Accordingly, there exists $t^*$ such that $\forall t \geq t^*, f(t) \leq g(t)$, and by algorithmic construction of $f \wedge g$ we can say that $T_{f \wedge g}$ is in fact such $t^*$.[6]

Therefore, we can apply Theorem 2, and compute $f \otimes g$ by:

- Computing $h = f \otimes g_a$. Since $\forall t \geq t^*, g_a(t) = +\infty$, computing this convolution will involve $d = d_f$ rather than $d = \mathrm{lcm}(d_f, d_g)$, thus smaller $D$ and $S_f^D$, reducing the cost of computation.
- Computing $f \otimes g = h \wedge f$. Being a minimum, it has a linear cost, but again $d = \mathrm{lcm}(d_h, d_f) = \mathrm{lcm}(d_f, d_f) = d_f$, hence the number of operations is greatly reduced.

The main benefit of applying Theorem 2 lies in dispensing with computing the representation of $f$ and $g$ over a possibly very long period $d = \mathrm{lcm}(d_f, d_g)$.

If neither of the above theorems can be applied, we can resort to the following property:

*Theorem 3 (Convolution of Sub-Additive Functions as Self-Convolution of the Minimum):* Let $f$ and $g$ be sub-additive functions $\in \mathcal{U}$ such that $f(0) = g(0) = 0$, and $f \otimes g$ is well defined. Then,

$$f \otimes g = (f \wedge g) \otimes (f \wedge g).$$

*Proof:* We recall that if $f$ is sub-additive with $f(0) = 0$, then [9, Corollary 3.1.1][7];

$$f \otimes f = f; \tag{16}$$

if $f(0) = g(0) = 0$, then [9, page 113][8]

$$f \wedge g \geq f \otimes g. \tag{17}$$

Then

$$(f \wedge g) \otimes (f \wedge g) = (f \otimes f) \wedge (f \otimes g) \wedge (g \otimes f) \wedge (g \otimes g)$$
$$\stackrel{(16)}{=} f \wedge (f \otimes g) \wedge (f \otimes g) \wedge g$$
$$= f \wedge g \wedge (f \otimes g)$$
$$= (f \wedge g) \wedge (f \otimes g)$$
$$\stackrel{(17)}{=} f \otimes g.$$

□

To exploit this theorem, we would first need to compute $f \wedge g$. However, this computation is also a prerequisite for testing Theorem 1, which one would try first anyway. Theorem 3 transforms a convolution into a *self-convolution*. Self-convolutions can be computed more efficiently than standard convolutions. In fact, we can bypass more than half of the elementary convolutions within the by-sequence algorithm, as per the following properties:

*Property 3 (Avoiding Duplicates in Self-Convolutions):* A self-convolution $h \otimes h$, $h \in \mathcal{U}$, can be computed through a single by-sequence convolution with $S_h^D \otimes S_h^D$, with $D = [0, 2 \cdot T_h + 2 \cdot d_h[$.

Since this by-sequence convolution is symmetric, we can reduce the number of its elementary convolutions to

$$\frac{n^2 - n}{2} < n^2,$$

where $n = N(S_h^D)$.

*Proof:* Since the two operands of the convolution have the same $\rho_h$, from [24] we know that:

- $T = T_h + T_h + d = 2 \cdot T_h + d$[9];
- $d = \mathrm{lcm}(d_h, d_h) = d_h$;
- $c = \rho \cdot d = c_h$.

Consider then $S_h^D, D = [0, T + d[ = [0, 2 \cdot T_h + 2 \cdot d_h[$, and its composing elements $e_i$, $1 \leq i \leq n$. It is:

$$S_h^D = e_0 \wedge e_1 \wedge \cdots \wedge e_n;$$
$$S_h^D \otimes S_h^D = \bigwedge_{e_i, e_j} e_i \otimes e_j.$$

The by-sequence convolution entails $n^2$ elementary convolutions. However, convolution being commutative, many of these are computed twice, e.g., $e_1 \otimes e_2$ and $e_2 \otimes e_1$. This can be avoided by computing instead:

$$S_h^D \otimes S_h^D = \bigwedge_{e_i, e_j : i, j \in [0..n-1]} e_i \otimes e_j$$
$$= \left( \bigwedge_{e_i, e_j : i, j \in [0..n-1], i < j} e_i \otimes e_j \right) \wedge$$
$$\left( \bigwedge_{e_i : i \in [0..n-1]} e_i \otimes e_i \right),$$

which results in $n + (n-1) + \cdots + 1 = \frac{n^2 - n}{2}$ elementary convolutions. □

Therefore, in a self-convolution (such as the one of Theorem 3) one can halve the number of elementary convolutions. On top of that, a further improvement is warranted by the following property:

*Property 4 (Reducing the Number of Convolutions by Element Coloring):* Let $f$ and $g$ be sub-additive functions $\in \mathcal{U}$ such that $f(0) = g(0) = 0$, and $h = f \wedge g$ (thus, $h(0) = 0$). Let $S_h^D$ be the sequence necessary to compute $h \otimes h$. Given *colors* $\{f, g\}$, for an element $e_k \in S_h^D$ defined in interval $D_k$, we define its color as

$$color(e_k) = \begin{cases} f & \text{if } \forall t \in D_k, e_k(t) = f(t) \\ g & \text{otherwise.} \end{cases}$$

---

[6]The alert reader may note that this is not true if *minimization of the transient* is applied to $f \wedge g$ – we indeed backup $T_{f \wedge g}$ beforehand.

[7]Once again, [9, Corollary 3.1.1] assumes wide-sense increasing functions. We leave to the interested reader the straightforward task to check that this hypothesis is not necessary to the correctness of the proof.

[8]See the previous footnote.

[9]When combined with Theorem 3, we can use $T = \min(2 \cdot T_{f \wedge g}, T_f + T_g)$.

An element's color is thus the function ($f$ or $g$) it belongs to. Then, we can compute $S_h^D \otimes S_h^D$ as:

$$S_h^D \otimes S_h^D = \left( \bigwedge_{\substack{e_i, e_j \in S_h^D \\ color(e_i) \neq color(e_j)}} e_i \otimes e_j \right) \wedge S_h^D, \quad (18)$$

i.e., we can omit computing elementary convolutions of elements of the same color.

*Proof:*

Since $h(0) + h(t) = h(t)$, we can write the convolution as

$$(h \otimes h)(t) = \inf_{0 \leq s \leq t} \{h(s) + h(t-s)\}$$
$$= \inf_{0 < s < t} \{h(s) + h(t-s)\} \wedge h(t).$$

Thus, we can ignore in the computation any pair $(t_i, t_j)$, such that $t_i + t_j = t$, for which $h(t_i) + h(t_j) \geq h(t)$. We show that elements of the same color fall in such category.

Let $e_i, e_j$ be elements of $S_h^D$ defined, respectively, on intervals $D_{e_i}, D_{e_j}$ and such that $color(e_i) = color(e_j) = f$. Let $D_{e_i \otimes e_j} = \{t = t_i + t_j \mid t_i \in D_{e_i}, t_j \in D_{e_j}\}$.

Then, for any $t \in D_{e_i \otimes e_j}$ and $t_i \in D_{e_i}, t_j \in D_{e_j}$ such that $t = t_i + t_j$, we have that:

$$(h \otimes h)(t) \leq h(t_i) + h(t_j)$$
$$= f(t_i) + f(t_j),$$

since $color(e_i) = color(e_j) = f$. On the other hand, due to sub-additivity of $f$,

$$f(t_i) + f(t_j) \geq f(t) \geq h(t).$$

Thus, $(e_i \otimes e_j)(t) \geq h(t)$. Obviously, the same holds if $color(e_i) = color(e_j) = g$.

Therefore, in order to compute $S_h^D \otimes S_h^D$, we only need to include in the computation of the lower envelope the sequence $S_h^D$ and the convolutions of elements with *different* colors, hence Equation (18).

$\square$

The idea behind Property 4 can be visualized through the example in Figure 18. Take $f$ and $g$ (Figure 18a), which intersect infinitely many times – hence their convolution cannot be simplified leveraging dominance. Figure 18b and Figure 18c report $S_{f \wedge g}^D$ against elementary convolutions $e_i \otimes e_j$, where $e_i$ and $e_j$ have the same color ($f$ and $g$, respectively). These figures show that the results of these elementary convolutions are always above $f \wedge g$. Instead, in Figure 18d we see how convolutions of elements of *different* colors may yield elements below $f \wedge g$.

The above two properties allow one to make the computation of $(f \wedge g) \otimes (f \wedge g)$ as efficient as possible, skipping many elementary convolutions. However, it remains to be seen whether computing the above is faster than computing $f \otimes g$ directly. Our results, reported in Section V-A4, show that this is indeed the case in the vast majority of cases: the ensuing time reduction ranges from sizeable percentages to 10 times. Counterintuitively, this is not due to a reduction in the number of elementary convolutions (which is instead of the same order of magnitude in the two cases, despite the optimizations of Property 3 and Property 4). Rather, it is due to the different topological properties of the ensuing elements. A thorough discussion of this phenomenon is reported in Section V-A4.

With reference to the example presented in Section III, we repeated the same computations, this time exploiting also the theorems proved in this section.

The new results are in Table VI and Table VII, which highlight further reductions in computation time.

## V. PERFORMANCE EVALUATION

In this section, we first evaluate the impact of our findings by measuring the speedup that they yield with respect to the standard algorithms described in Section II-A, taken from [25]. Then, we show how our method allows one to analyze long tandems of flow-controlled nodes, and we compare the exact and approximate analysis methods as for efficiency and accuracy.

### A. Computational Results

We run our experiments on a desktop PC equipped with an Intel Core i9-9900, 16 GB of DRAM @3200 MHz, Windows 10. Our publicly available NC library, called Nancy [30], [31] is written in C# (.NET 6), and can exploit the PLINQ framework to easily parallelize and scale the algorithms in multicore processors. However, to minimize the perturbations and improve the fairness of comparisons, our experiments are run in single-thread mode. This allows us to obtain consistent time measurements: we verified that the execution times of independent replicas of the same experiment differ by fractions of percentage points. For this reason, confidence intervals are omitted. As numerical base type we introduce a *Rational* type, in which both numerator and denominator are represented using *System.Numerics.BigInteger*, which is an integer type with no upper or lower bounds. This comes at a performance cost over using 64-bit integers, but has the distinctive advantage of removing all issues with arithmetic overflow. Execution times are measured using the *System.Diagnostic.Stopwatch* class. When applying our optimizations, the execution times we measure also include those spent testing our hypotheses (e.g., dominance or asymptotic dominance).

In the experiments, we focus on convolutions between UPP curves in the form $\beta_{R,\theta,h} = \overline{\beta_{R,\theta} + h}$, where $\beta_{R,\theta}$ is a rate-latency curve, with latency $\theta$ and rate $R$, and $h$ is the ordinate of a constant function.

Parameters $R, \theta, h$ are generated using a pseudo-random generator (*System.Random*) that produces integers between 1 and 1000. The resulting curves are further filtered in order to match the properties required by the theorems being tested. We analyze separately the speedup obtained with representation minimization and with the Theorems described in Section IV.

*1) Representation Minimization:* We now test the impact of representation minimization, described in Section IV-A. We compute $(\beta_a \otimes \beta_b) \otimes \beta_c$, where the three operands are randomly generated as described above. First, we computed the convolutions without any improvement. Then, we computed the same convolutions using representation minimization both on the results and in between any intermediate step: for instance, when we compute Section II-A, we minimize the result of each of the four partial convolutions. Note that the algebraic properties of Theorems 1 to 3 were not used in this study.

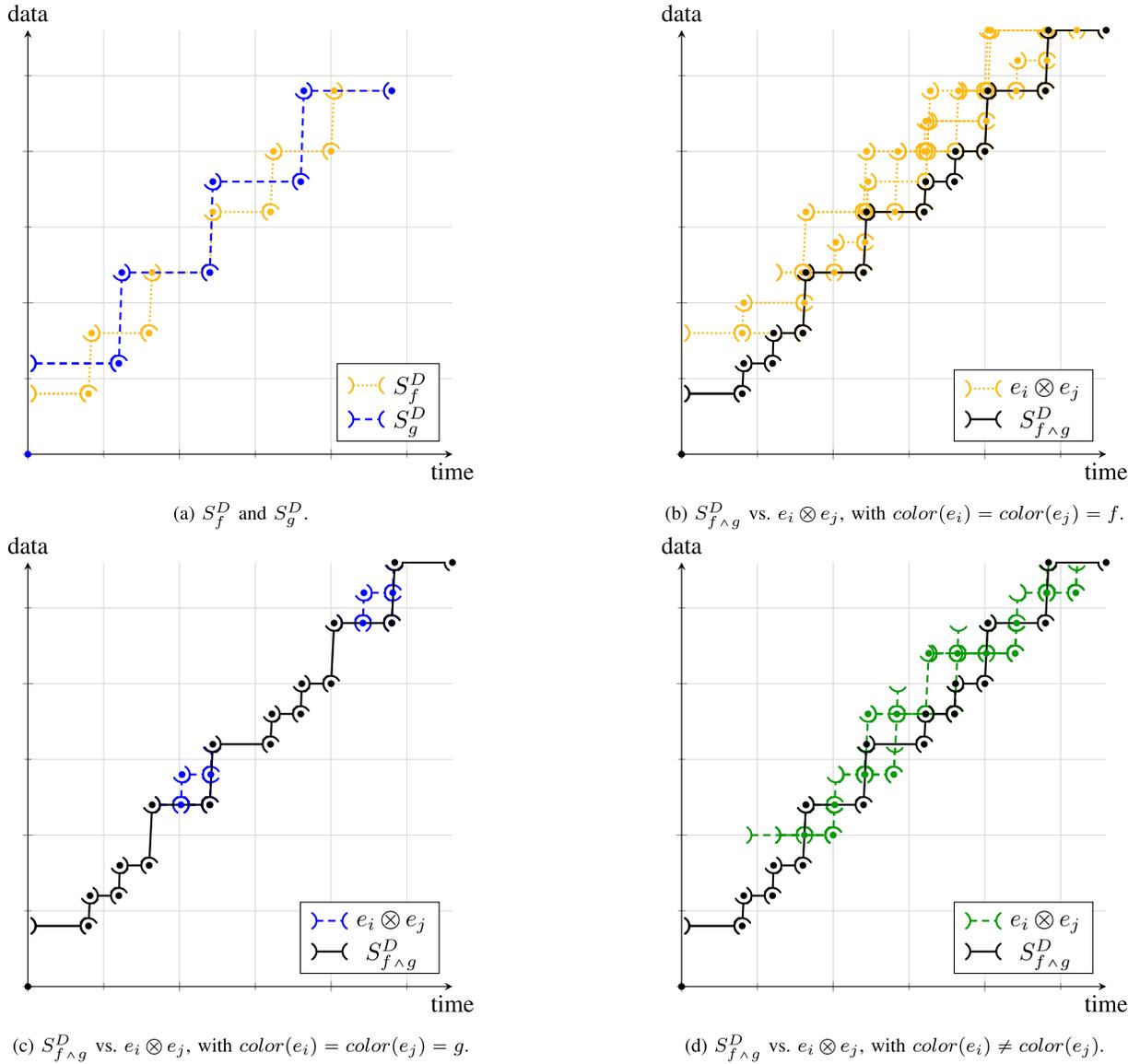

Fig. 18. Coloring example.

TABLE VI
COMPUTATIONAL RESULTS, EXACT METHOD

|  | w/o optimizations | minimization | minimiz. + Th. 1,2,3 |
|---|---|---|---|
| comp. time of $\overline{\beta_2 \otimes \beta_3^{eq} + W_3}$ | 6 h 24 m | 6 s | 0.47 s |
| $N(\beta_2 \otimes \beta_3^{eq} + W_3) \to N(\overline{\beta_2 \otimes \beta_3^{eq} + W_3})$ | $10 \to 10600$ | $10 \to 10$ | $10 \to 10$ |
| comp. time of $\overline{\beta_1 \otimes \beta_2^{eq} + W_2}$ | > 24 h | 13 s | 0.18 s |
| $N(\beta_1 \otimes \beta_2^{eq} + W_2) \to N(\overline{\beta_1 \otimes \beta_2^{eq} + W_2})$ | unknown | $14 \to 6$ | $14 \to 6$ |

TABLE VII
COMPUTATIONAL RESULTS, APPROXIMATE METHOD

|  | w/o optimizations | minimization | minimiz. + Th. 1,2,3 |
|---|---|---|---|
| comp. time of $\beta_{\{1,3\}} = \overline{\beta_1 \otimes \beta_2 + W_2} \otimes \overline{\beta_2 \otimes \beta_3 + W_3}$ | 0.14 s | 0.11 s | 0.09 s |
| $N\left(\overline{\beta_1 \otimes \beta_2 + W_2}\right), N\left(\overline{\beta_2 \otimes \beta_3 + W_3}\right) \to N\left(\beta_{\{1,3\}}\right)$ | $6, 6 \to 270$ | $6, 6 \to 42$ | $6, 6 \to 42$ |
| comp. time of $\beta_{\{1,4\}} = \beta_{\{1,3\}} \otimes \overline{\beta_3 \otimes \beta_4 + W_4}$ | 6 h 13 m | 13.47 s | 0.003 s |
| $N(\beta_{\{1,3\}}), N\left(\overline{\beta_3 \otimes \beta_4 + W_4}\right) \to N(\beta_{\{1,4\}})$ | $270, 6 \to 1456$ | $42, 6 \to 6$ | $42, 6 \to 6$ |

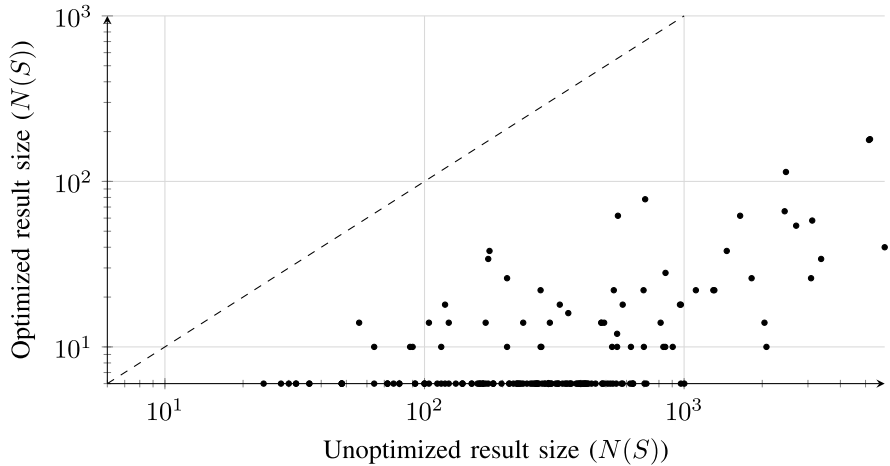

Fig. 19. Cardinality of the results of the convolution of three sub-additive functions, with and without minimization.

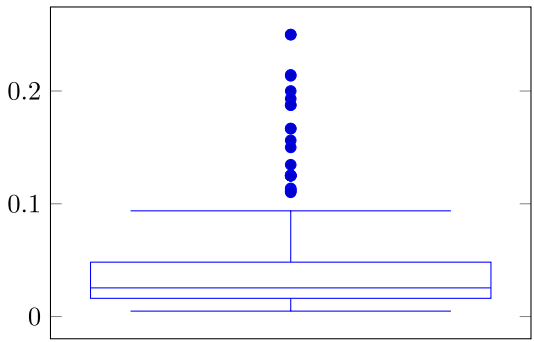

Fig. 20. Convolution of three sub-additive functions: reduction of the cardinality of the result due to minimization.

We first report in Figure 19 the reduction in the number of elements of the result. Each experiment is reported as a point on a Cartesian plane (with logarithmic scales): its ordinate is the number of elements of the minimized result, and its abscissa is the number of elements of the unoptimized one. The dashed line is the bisector, below which the representation reduction factor is larger than 1. This makes it easier to visualize the order of magnitude of the reduction, which is in fact the horizontal (or vertical) distance between a point and the bisector. Figure 19 highlights representation reductions of one to three orders of magnitude. A box plot of the representation reduction is shown in Figure 20.

We have presented in Section IV-A an example where representation minimization also yielded a speedup of orders of magnitude. We therefore evaluate the cost of the above operations in Figure 21, which compares the optimized and unoptimized running times, and in Figure 22, which reports a box plot of the reduction in computation times.

Our results show that – while some gains are certainly there in most cases – a high reduction in the state occupancy does not always yield a similar reduction in computation times. The median time reduction is in the order of $10\%$. This can be explained by observing that the impact of *period minimization* on the lcm is variable. Consider for example two curves, $f$ and $g$, and their respective representations with $d_f = 30$ and $d_g = 30$, such that by performing period minimization on these representations we obtain $\tilde{d}_f = 5, \tilde{d}_g = 6$. While the reduction in size is noticeable, the same cannot be said about computing $f \otimes g$, since $\text{lcm}(30, 30) = \text{lcm}(5, 6)$. Thus, the more substantial speedups are obtained when minimization succeeds in removing a common factor from both operands (e.g., factor 5 from both $f$ and $g$). Note that these cases (to which the example of Section IV-A belongs) depend on numerical properties of the operands, hence are hard to obtain using random generation of the input (but not impossible – check the outliers at the bottom of Figure 22).

We stress that the most remarkable benefits of representation minimization lie in enabling the computation of SACs, hence the analysis of networks via the exact method. In this case, representation minimization is indispensable, since the complexity of the SAC algorithm is exponential with the number of elements. Unfortunately, we are not able to produce a speedup figure for the SACs, since unoptimized SACs with random parameters hardly ever terminate at all.

On the other hand, the above experiments highlight that applying minimization always yields a significant size reduction, which helps with memory management; it yields at least a moderate time reduction, most of the times, and – even in the rare cases when it fails to provide a speedup – the time spent on applying it is negligible.

*2) Convolution of Sub-Additive Functions With Dominance:* We now test the impact of Theorem 1. We compute $\beta_{R_1,\theta_1,h_1} \otimes \beta_{R_2,\theta_2,h_2}$, where the operands are randomly generated and matching the hypotheses of Theorem 1. To make the comparison more insightful, in these and the following experiments we apply representation minimization to all intermediate computations in the baseline unoptimized algorithm.

The benefits of using Theorem 1 can be seen in Figure 23, which clearly shows that most speedups are in the region of $10^5$ times. In many cases the unoptimized convolution lasted more than 10 minutes, while the optimized version seldom lasted more than 1 ms. This means that dominance is a property worth checking.

*3) Convolution of Sub-Additive Functions With Asymptotic Dominance:* We compute $\beta_{R_1,\theta_1,h_1} \otimes \beta_{R_2,\theta_2,h_2}$, where the operands are randomly generated and matching the hypotheses of Theorem 2. The impact of Theorem 2 is shown in Figure 24, which still highlights speedups in the order of $10^5$ times: unoptimized convolutions taking several minutes are often reduced to fractions of a second. However, some of the lengthy computations still take a sizable time even after the optimization. This is because the effect of Theorem 2 is to use the time of last intersection, rather than $\text{lcm}(d_f, d_g)$, to determine the sequences to be convolved. In few cases, the former may exceed the latter, hence Theorem 2 may instead increase the cost (see the point above the bisector in the bottom-left corner of Figure 24). However, such cases are rare – and easy to avoid. In fact, we can compare the extremes of the extended intervals of the operands computed with the standard algorithm and Theorem 2 and then run the algorithm that will involve fewer elementary convolutions.

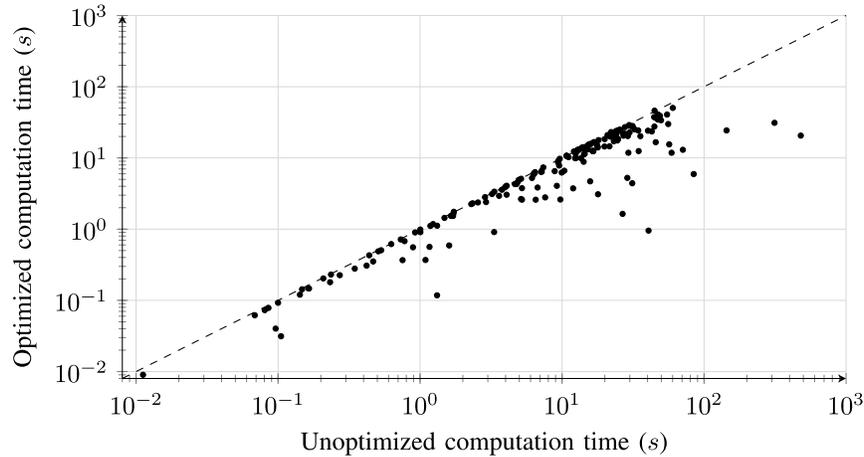

Fig. 21. Computation times of the convolution of three sub-additive functions, with and without minimization.

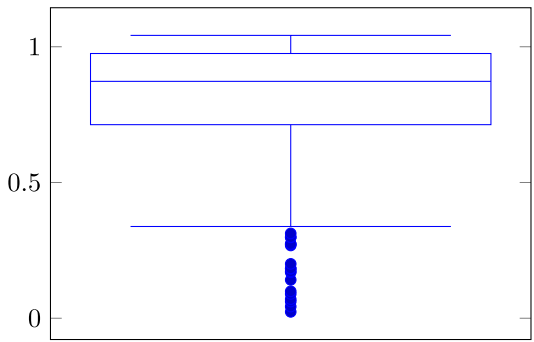

Fig. 22. Convolution of three sub-additive functions: reduction of computation times due to minimization.

*4) Convolution of Sub-Additive Functions as Self-Convolution of the Minimum:* A first assessment of the impact of Theorem 3 (coupled with Properties 3 and 4) is reported in Figure 25. It is evident from the figure that the speedup is less prominent in this case – the maximum that we get is 30 times. Note that the higher speedups are obtained when the unoptimized computations take more time (see the top-right cluster of points). However, there is a speedup in almost all cases – we only found one outlier at 0.99 times, meaning that using our theorem takes a little more time than using the basic convolution algorithm. The obtained speedup is mostly within one order of magnitude. For this reason, we report in Figure 26 a box plot of the *reduction of computation times* (which is the inverse of the speedup). With our method, computation times can be expected to be 30% to 80% of the unoptimized times.

Intuitively, one might expect the above speedup to be related to the number of elementary convolutions. However, Figure 27 shows that this is not the case: the number of elementary convolutions is roughly the same, regardless of the achieved speedup. In more than a few cases, applying Theorem 3 entails computing *more* elementary convolutions (i.e., all the points having abscissa smaller than 1), yet the optimized version yields a nonnegligible speedup nonetheless.

The root cause of the speedups lies elsewhere. To explain it, we first need to give some details about the by-sequence convolution algorithm, namely its final step, i.e., computing the lower envelope of the elementary convolutions.

In a convolution $f \otimes g$, one must compute the elementary convolution of each element in $S_f^D$ with each element of $S_g^D$ ($D$ being the extended interval, as shown in Theorem 8 in Appendix A). This yields a set $E$ of points and segments, whose lower envelope represents the resulting sequence. The topological relationship between any two elements $x, y \in E$ is unknown a priori. They may or may not intersect, they may belong to disjoint time intervals, one may dominate the other. This is exemplified in Figure 28.

In order to compute the lower envelope of $E$, the first step is to determine the *intervals of overlap*, henceforth *intervals* for short. An interval $I$ consists of a domain $D_I$, equal either to $[t_I, t_I]$ (point-sized) or $]t_I^s, t_I^e[$ (segment-sized), and a set of elements $E_I \subseteq E$ which are all the elements of $E$ that are defined $\forall t \in D_I$. The lower envelope is computed as the juxtaposition of the lower envelopes interval by interval. To compute and populate intervals, we:

- collect the start and end times of all the elements in $E$ and order them. Let the result be $t_0, t_1, \ldots, t_N$;
- derive point- and segment-sized domains for and between each of these times, i.e., $[t_0, t_0], ]t_0, t_1[, [t_1, t_1], \ldots$;
- from the above domains, derive the set of intervals, initialized with empty lists of elements;
- for each element $e \in E$, find all the intervals $e$ belongs to, and add $e$ to their lists.

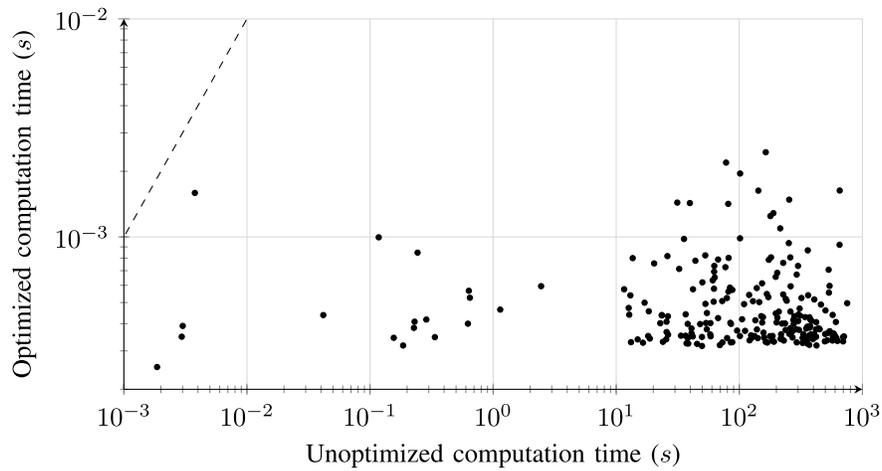

Fig. 23. Results of the convolution of sub-additive functions with dominance.

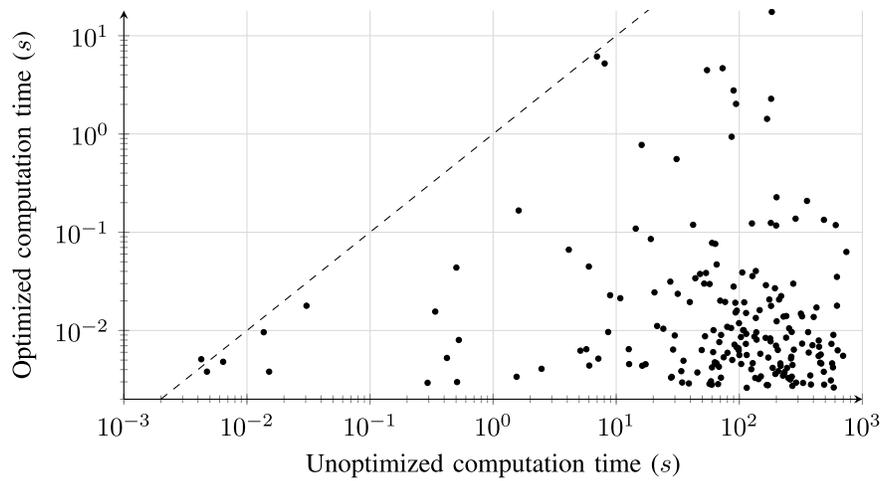

Fig. 24. Results of the convolution between sub-additive functions with asymptotic dominance.

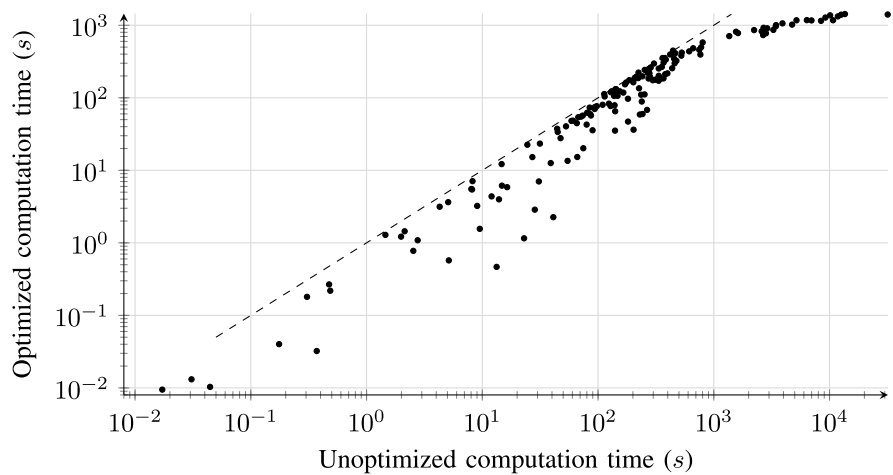

Fig. 25. Results of the convolution between sub-additive functions without asymptotic dominance.

We underline that the element-interval relationship is many-to-many: the same element may span multiple intervals, and an interval may include several elements. Afterwards, we compute the lower envelope of each interval $I$, and we concatenate them to obtain the overall lower envelope of $E$.

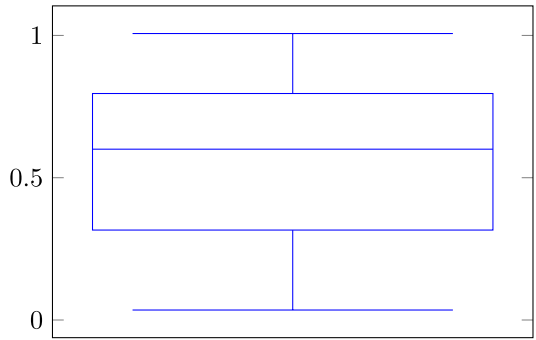

Fig. 26. Reduction of computation times of the convolution between sub-additive functions without asymptotic dominance.

TABLE VIII
PARAMETERS OF THE EXAMPLE CONVOLUTION

|   | $R$ | $\theta$ | $h$ |
|---|-----|----------|-----|
| $f$ | 901 | 499 | 192 |
| $g$ | 806 | 36 | $\frac{6912}{499}$ |

As for the algorithm costs, we note that:
1) *finding* which intervals an element belongs to is $\mathcal{O}(n \cdot \log(n))$, where $n$ is the number of intervals, if one uses an interval tree;
2) *inserting* an element in the lists of *all the intervals* it belongs to, instead, depends on the number of intervals an element belongs to (something which we show to be highly variable in a few lines), and is $\mathcal{O}(n)$ in a worst case;
3) computing the per-interval lower envelope of $I$ costs $\mathcal{O}(m)$ if $I$ is point-sized, $\mathcal{O}(m \cdot \log(m))$ if segment-sized, where $m$ is the cardinality of $E_I$;
4) the concatenation of the per-interval results is $\mathcal{O}(n)$.

Of the above steps, we note that steps 2 and 3 are independent of the number of elements, but instead depend on how much overlap there is between them. In fact, the more overlap there is between the elements of $E$, the higher the cost of this algorithm is. Some of the overlaps – actually, most – will not yield segments that end up being part of the lower envelope.

We show through a relevant example that computing $(f \wedge g) \otimes (f \wedge g)$ yields considerably less populated intervals than computing $f \otimes g$. The parameters are as in Table VIII.

In the non-optimized convolution algorithm, we need to compute the lower envelope of 810k elements, for which 220k intervals are used. In the optimized algorithm, we find instead 910k elements and 320k intervals. However, as Figure 29 highlights, there is a significant difference in how many intervals each element spans. This affects the cost of step 2, which takes $180s$ in the non-optimized algorithm vs. $42s$ in the optimized one. Moreover, as Figure 30 highlights, there is also a significant difference in how many elements a given interval list includes, which affects the cost of computing the per-interval lower-envelope. In fact, step 3 takes $370s$ in the non-optimized algorithm, against $70s$ in the optimized one.

Overall, applying the optimizations discussed produces, in this example, a fivefold speedup – which is counter-intuitive if one considers only the number of convolutions.

### B. A Case Study

We now show how our method allows one to analyze flow-controlled networks. We consider a tandem of $n$ nodes, $n = 2 \ldots 10$, where all nodes are described by the same rate-latency service curve $\beta_{16,2}$, and with input buffers of increasing size $W = 13, 15, \ldots, 29$. In Figure 31 we compare the running times of the exact method (10) and the approximate method (12), with and without the optimizations described in this paper. We observe that the exact method can only be run *with* our optimizations: without them, the computations for a three-node tandem had not completed after 24 hours. The graph clearly shows that the approximate method is orders of magnitude faster than even the optimized exact one. However, our optimizations still take away one order of magnitude of computations in that as well. The experiments were run five times in independent conditions, and 95% confidence intervals were always within 1% of the average. For that reason, they are not reported in the graph.

We found that the computation times (whichever the method) are very sensitive to the actual parameters of the network: changing the numbers in the above example is likely to change the vertical scale of the above graph considerably. However, the same pattern still emerges: the unoptimized exact method is just unfeasible most of the times; the optimized exact method comes second; the approximate method is considerably faster, and even faster with our optimizations.

To support the above claim, we present another scenario in Figure 32, where the computation times for the approximate method are sensibly higher. To obtain such a difference, all it took was to modify rates to $R = 1600$, latencies to $\theta = 200$, and buffer sizes to $W = 1300, 1305, \ldots, 1340$. In this case, the approximate method takes up to hundreds of seconds, whereas our optimizations curb the computations at fractions of a second. It is interesting to observe that our optimization yield times that are non monotonic with the tandem length (see, e.g., around $n = 8$). This is because a more favorable optimization kicks in at $n = 8$ and further abates computation times.

What our optimizations allow – for the first time, to the best of our knowledge – is an assessment of the *accuracy* of the approximate method. In fact, this requires being able to complete exact computations, which just cannot be done without these very optimizations (unless one handpicks very trivial scenarios and parameter values, with the obvious risk of undermining generality). Our results here are quite surprising. They show that the end-to-end service curves obtained via the approximate method are *always equal* to the exact ones. This occurs not only in the tandems described in this paper, but in all the cases we analyzed, including many (several tens) with randomized configurations.

One may legitimately wonder if this is due to the fact that equality should hold in Equation (13), but so far no one was able to prove it. We show that this is not the case, i.e., there are cases when $\beta_i^{eq} > \beta_i^{eq'}$. Consider the three-node tandem in Figure 7, and assume that nodes have the same rate-latency service curve $\beta_{16,2}$, and with input buffers $W_2 = 20$ and $W_3 = 13$.

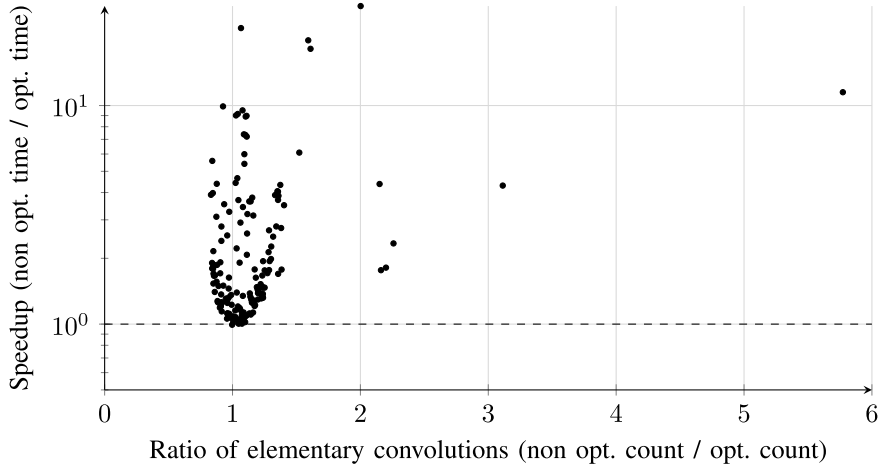

Fig. 27. Ratio of elementary convolutions vs. speedup.

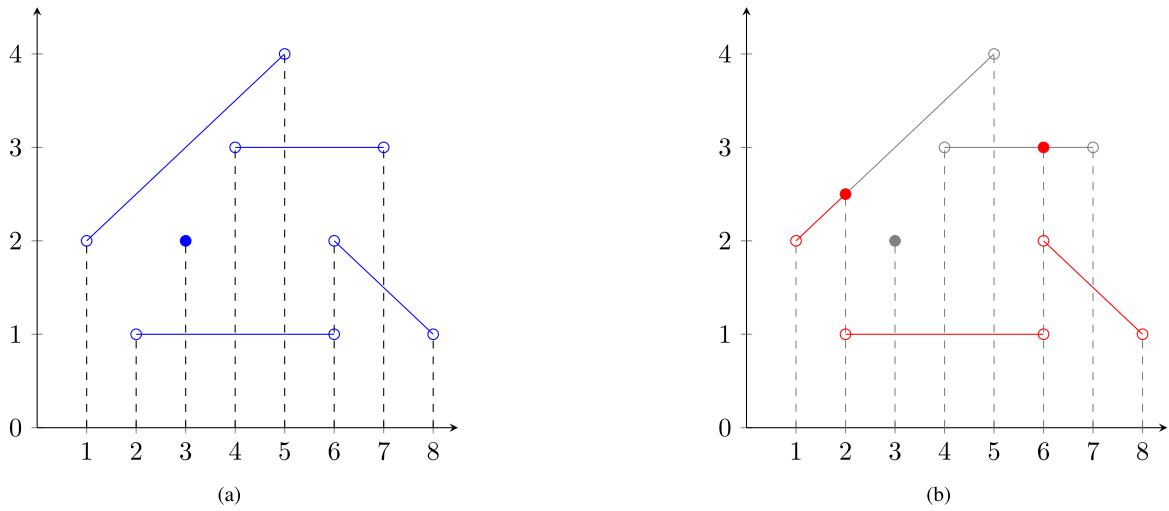

Fig. 28. Example of intervals for a set of elements (a) and their lower envelope (b).

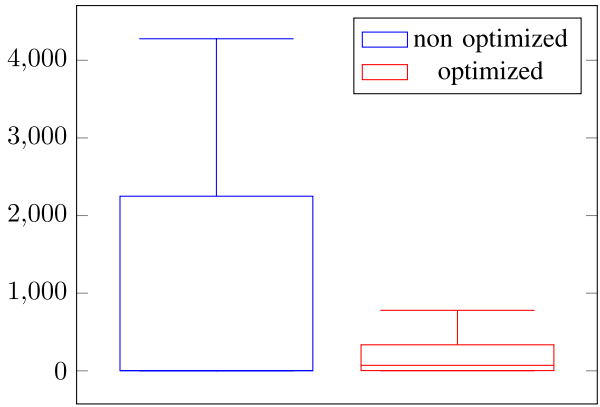

Fig. 29. Number of intervals per element.

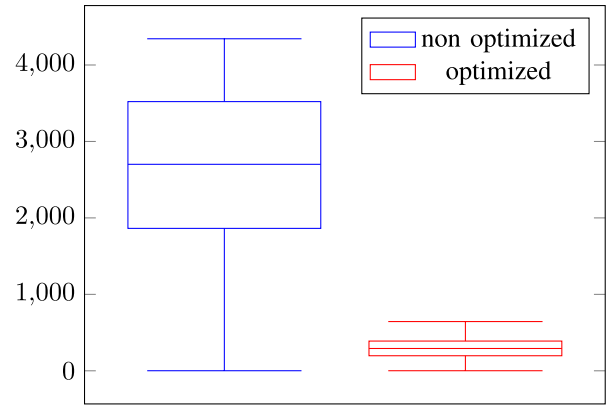

Fig. 30. Number of elements per interval.

When computing the equivalent service curve at the *first* node, i.e., $\beta_1^{eq}$, $\beta_1^{eq'}$, we obtain different results using the exact and approximate method, as shown in Figure 33a. It is $\beta_1^{eq} > \beta_1^{eq'}$. The difference can be explained by observing that, since $W_2 > W_3$, it is expected that the worst-case performance will be initially constrained by the larger buffer $W_2$ (see the first step in Figure 33a), then by the smaller buffer downstream (see the second step onwards in the same figure). The exact computation reflects this phenomenon, while the approximate method does not. However, despite this, Figure 33b shows that this difference is irrelevant when computing the equivalent service curve for the whole tandem. It is in fact $\beta^{eq} = \beta^{eq'}$. A similar phenomenon was observed in all our experiments.

The above observations cast the approximate method in a new – and more favorable light. They suggest that the latter is as performing as the exact one, in an end-to-end context. This is important, because one can always find cases where – despite our optimizations – the exact method will just be too costly.

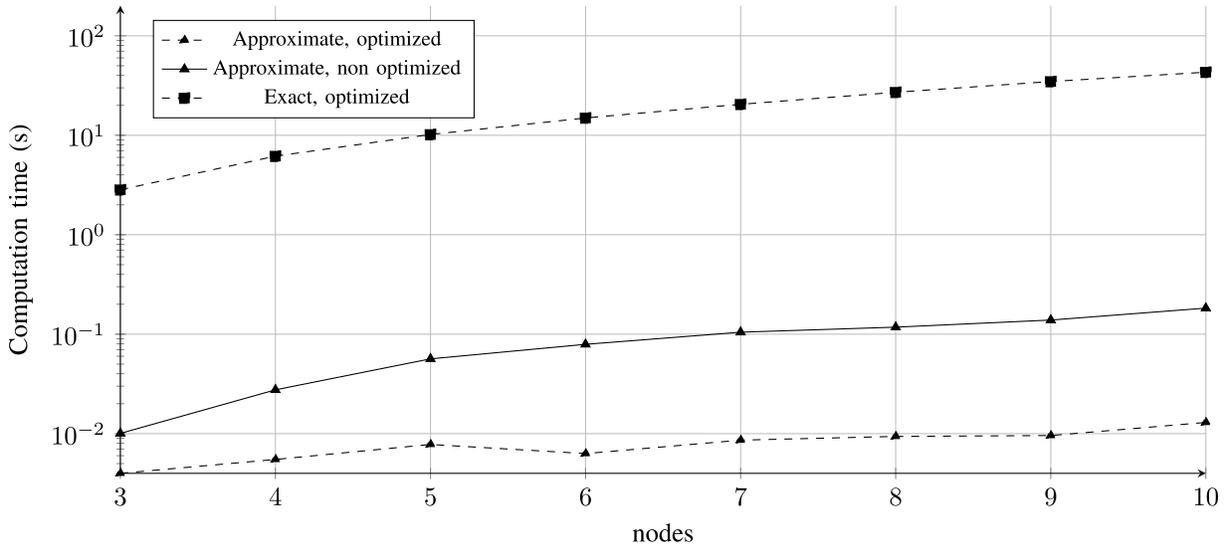

Fig. 31. Performance comparison of the exact and approximate methods.

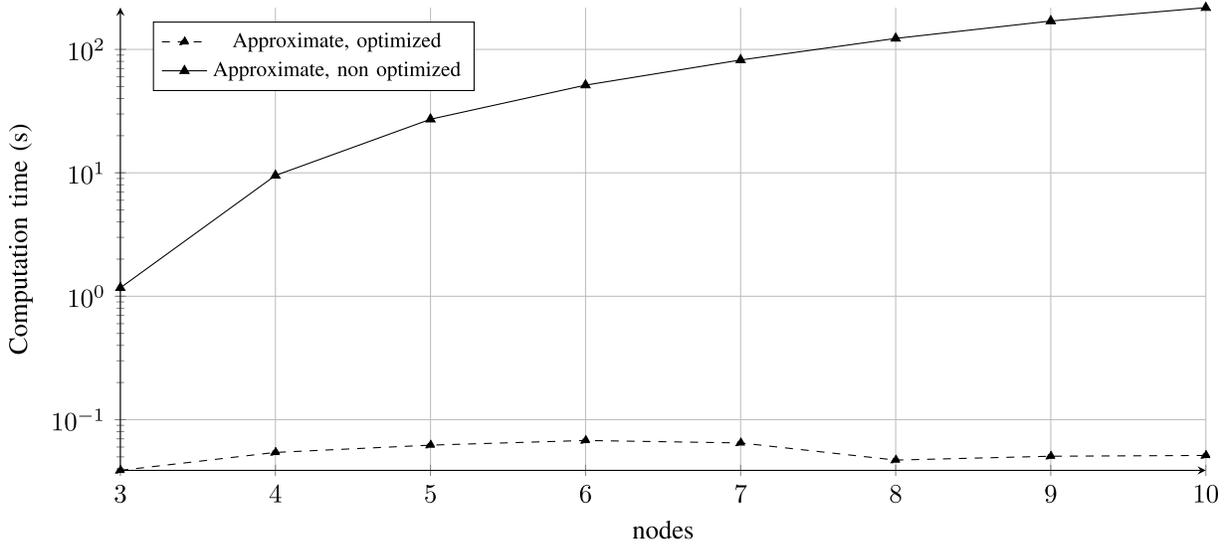

Fig. 32. Performance comparison of the optimized/unoptimized approximate method.

## VI. Related Works

The theory of Network Calculus dates back to the early 1990s, and it is mainly due to the work of Cruz [7], [8], Le Boudec and Thiran [9], and Chang [10]. Since then, a considerable number of papers have extended it to include different scheduling algorithms and flow multiplexing schemes [32], [33], [34], [35], extensions to stochastic characterizations of service and traffic [36], [37], [38], specific network architectures [4], [5], [15], [19], [26], [39].

The *computational* aspects of NC implementations have been the subject of several papers in the past. Problems such as efficient data structures to represent arrival/service curves or functions, or the complexity of NC operators (e.g., convolution or sub-additive closure) have been tackled in the works of Bouillard et al. [24], and find a thorough exposition in book [25], which also reviews the implementations of several existing tools. The idea of UPP curves as a class closed with respect to NC operations is in fact reported in these works. We use the NC algorithms described therein as a baseline. A related research field is that of Real-time Calculus (RTC), developed for real-time systems [40]. RTC is based on min-plus and max-plus operators that work on Variability Characterization Curves, which appear to be very similar to UPP functions (although [24] observes that the two classes treat discontinuities in a different way, and [41] remarks that they were never formally compared). RTC uses min-plus convolution to obtain the output of a system given its input, similarly to NC. For this reason, the two methodologies have often evolved via cross-fertilization, with solutions devised in one context often being ported to the other. In fact, RTC

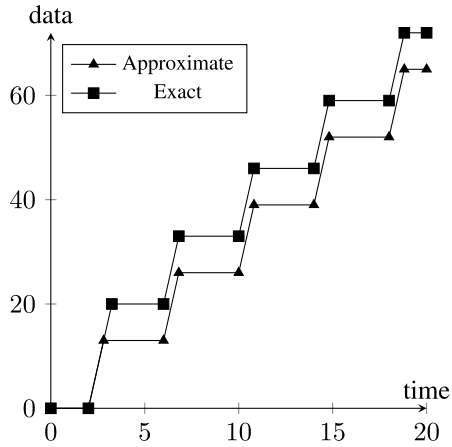

(a) Comparison of equivalent service curves at node 1 $\beta_1^{eq}$, $\beta_1^{eq'}$.

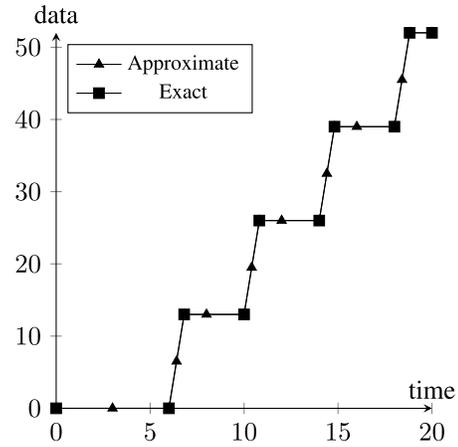

(b) Comparison of equivalent end-to-end service curves of the tandem $\beta^{eq}$, $\beta^{eq'}$.

Fig. 33. Comparison of the results of the exact and approximate method.

work [27] first observed that multi-hop traversal – which entails chained convolutions – is subject to state explosion, and that the latter makes convolutions exponentially complex. They proposed a way to mitigate this problem, which relies on inferring the maximum time beyond which the shape of the resulting functions is immaterial, which turns out to be considerably smaller than the lcm of the periods, thus leading to more efficient operations. This idea of a compact domain is transferred to NC in [28] – allowing it to be used in conjunction with NC service curves. Work [42] further generalizes it to more operations and more general RTC settings. NC analysis limited to compact domains [28] consists in finding finite upper bounds to the time where operations should be computed. This allows by-sequence operations to be computed between two finite sets of elements – which one can imagine as transient parts – disregarding periodicity and the lcm explosion that comes with it. The upper bound is chosen so that the end-to-end delay and backlog analysis is not affected. This is done by working with lower/upper approximations of *both* the arrival curve and the service curves, using concave/convex piecewise linear (CPL) curves, which is computationally inexpensive. As already explained, this method cannot be applied to our problem, since it leverages *super-additivity* of service curves, a property that does not hold in our settings.

As far as NC *tools* are concerned, a critical review of the available software packages is reported in [43]. However, the latter only reports *functional* comparisons (i.e., discusses the capabilities of each tool), and does not address performance issues. To the best of our knowledge, there seems to be no other *open* NC tool that is able to deal with UPP functions, beside ours [30], [31]. Most of the existing public tools, e.g., CyNC [44], NC-Maude [45], DiscoDNC [46], DEBORAH [47], [48], CATS [49] restrict the implementation of NC operators to the case of CPL functions. This means that they cannot be used to run the computations described in this paper. The COINC library [50] did address UPP (implementing the algorithms in [24]), but it appears that it is no longer available. Commercial tool Real-Time-at-Work (RTaW) can perform min-plus computations on UPP functions, and is also available via a browser-based interpreter [51]. However, its license explicitly prohibits using it for benchmarking purposes. The RTC algorithmic toolbox [52] is a Java-based tool for RTC analysis, available also as a Matlab toolbox, which implements a general-purpose convolution operator, that should be able to run – in theory – the examples of Section V. To the best of our knowledge, RTC's source code is not available at the time of writing this paper. We have no indication that it uses our findings in its computations. However, RTC uses floating-point arithmetics, which means that it may be subject to numerical errors whose impact is difficult to assess. Our arithmetics is based instead on rational numbers, hence computations are always exact. Moreover, it seems that RTC loops infinitely in several cases – which are not challenging, performance-wise [30].

Systems with flow control have traditionally been analyzed using Markov Chains [6], under the name of "queueing systems with blocking". That method allows one to find mean performance indexes (and, possibly, distributions), starting from a *stochastic* characterization of input traffic and service. The first works analyzing flow control in the framework of NC have been [53], [54]. The exact method – i.e., the one using nested SACs – is a direct application of these results. The approximate method – i.e., the one using convolution of SACs – is instead shown in [26]. This paper, however, does not assess the gain in efficiency warranted by the approximate method, nor it acknowledges the fact that it seems to preserve accuracy. We argue that this may be due to the fact that the computational problems addressed in this paper were in the way of such an evaluation. A different use case with hop-by-hop flow control is studied in [29], which focuses on Stream Processing Systems in Real-Time Calculus. It is shown therein [29, Theorem 3] that an effective service curve for the first node in a tandem can be computed via a chain of convolutions of sub-additive expressions, i.e., the same type whose computation we optimize in this paper. Paper [55] uses a NC model with flow control to model Denial-of-Service (DOS) attacks. Flow control is also addressed in [56], in the framework of *stochastic* NC.

## VII. CONCLUSION AND FUTURE WORK

In this paper, we addressed the problem of analyzing tandems of flow-controlled network elements. We reviewed the available methods, both exact and approximate, and highlighted that both can be onerous from a computation standpoint. The exact method, in particular, scales rather poorly with the tandem length, virtually making analysis infeasible if the tandem has three or more hops. We showed that the approximate method – although computationally more affordable – may still scale poorly. We traced back the problem to the explosion of hyper-periods in the convolution of ultimately pseudo-periodic curves, a problem which cannot be mitigated using known techniques (which rely on hypotheses that our equivalent service curves fail to verify). We then presented novel computational and algebraic techniques to reduce the computation time of convolutions. Our first technique consists in minimizing the representation of a function after every operation. In fact, the existing algorithms for NC operations yield non-minimal representations, and the length of a representation is the dominating factor in the complexity of the algorithms. We showed that minimization is computationally cheap. It may yield reductions in the representation size of orders of magnitude. This translates to a similar reduction in the *number* of elementary convolutions involved in a SAC, hence acting as an enabler for exact analysis. Moreover, by reducing the length of the periods, it may also reduce the lcm of the periods of the operands in a convolution, thus making it more efficient. On top of that, we presented novel algebraic properties of sub-additive functions that lead to optimized convolution algorithms. More specifically, we showed that the convolution of sub-additive functions can be greatly simplified if a *dominance* relationship exists between the two: it is either a simple minimum, or a convolution of generally shorter sequences, hence much faster. Moreover, even when dominance cannot be leveraged, we can always transform the convolution of sub-additive functions into a self-convolution of their minimum. Self-convolution of a minimum is a rather efficient operation, since – on one hand – it allows one to filter away several elementary convolutions, thus reducing the computation time, and – on the other – it yields sequence elements that have better topological properties, making the final computation of a lower envelope considerably more efficient. We have showed that the speedup brought by our algebraic properties ranges from two-digit percentages (especially in the case of self-convolution of the minimum) to several orders of magnitude (in the other cases), even factoring in the time required to check the properties themselves. The cases when little or nothing is gained in the way of efficiency are a small minority, and can often be identified *a priori*. Our optimizations allowed us to compare the exact and approximate analysis methods, as for efficiency and accuracy, something that could not be done before due to the prohibitive times involved in the computations. Our results are that – rather surprisingly – the approximate method seems to be as accurate as the exact one, because differences in the per-node equivalent service curves get erased in an end-to-end context. This is particularly important, since the approximate method scales considerably better than the exact one.

We observe that the techniques outlined in this paper may lend themselves to other applications. For instance, representation minimization can always be applied when performing NC operations on UPP curves. Sub-additive UPP curves, moreover, may come up for other reasons than the curve being the result of a SAC.

There are several directions in which our work can be expanded. On the algebraic side, looking for other classes of functions (beside sub-additive ones) for which similar properties as those shown in this paper hold, possibly leading to similar simplifications. Moreover, the fact that – in all the experiments we performed – the exact and approximate method end up computing the same end-to-end service curve from different per-node service curves, clearly calls for further investigation. On the computational side, we can observe that several of the complex tasks to be performed with operations (e.g., finding the lower envelope interval by interval) are amenable to parallel implementation. Our NC library already supports parallelization, thus our next endeavor is to investigate this avenue further, to understand what can be gained by distributing what tasks among parallel threads.

## APPENDIX A
## MIN-PLUS OPERATIONS ON UPP CURVES

We report below the proofs of the algorithms to compute the results of min-plus operations for functions in $\mathcal{U}$, i.e., ultimately pseudo-periodic piecewise affine $\mathbb{Q}_+ \rightarrow \mathbb{Q} \cup \{+\infty, -\infty\}$ functions. Proofs are adapted from those in [24], with a few clarifications of our own. The *stability* of min-plus operators for the above class of functions is discussed in [24]. For the sake of conciseness, we will not discuss here how to compute elementary operations, i.e., between points, segments and limited piecewise sequences.

We recall that, for functions in $\mathcal{U}$,
- $\forall t \geq T, f(t + k \cdot d) = f(t) + k \cdot c, k \in \mathbb{N}$;
- $\rho = c/d$.

*Minimum*

For the minimum $f_1 \wedge f_2$, we need to treat the following two cases separately:
- $\rho_1 = \rho_2$;
- $\rho_1 < \rho_2$ (without loss of generality, minimum being commutative).

*Theorem 4 (Minimum of Pseudo-Periodic Functions With the Same Rate):* If $\rho_1 = \rho_2 := \rho$, $f_1 \wedge f_2$ is pseudo-periodic with $T = \max(T_1, T_2)$, $d = \text{lcm}(d_1, d_2)$, $c = \rho \cdot d$.

*Proof:*

$\forall t \geq \max(T_1, T_2)$
$$(f_1 \wedge f_2)(t + d) = f_1(t + d) \wedge f_2(t + d)$$
$$= \left(f_1(t) + \frac{d}{d_1} \cdot c_1\right) \wedge \left(f_2(t) + \frac{d}{d_2} \cdot c_2\right)$$
$$= (f_1(t) + \rho \cdot d) \wedge (f_2(t) + \rho \cdot d)$$
$$= f_1(t) \wedge f_2(t) + \rho \cdot d.$$

We observe that, in order to be able to leverage the pseudo-periodicity property for both $f_1$ and $f_2$ in this proof:
- it is enough that $t$ is the largest between $T_1$ and $T_2$, i.e., $t = \max(T_1, T_2)$;
- it is necessary that $d = \text{lcm}(d_1, d_2)$, as both $\frac{d}{d_1}$ and $\frac{d}{d_2}$ need to be $\in \mathbb{N}$.

$\square$

*Theorem 5 (Minimum of Pseudo-Periodic Functions With Different Rates):* If $\rho_1 < \rho_2$, $\exists \; \bar{t} : \; \forall t \geq \bar{t}, f_1(t) \leq f_2(t)$. In particular, we can compute $\bar{t}$ as follows:
- define the upper boundary of the pseudo-periodic behavior of $f_1$ as the line $U_1(t)$ such that $\forall t \geq T_1, f_1(t) \leq U_1(t)$. We can compute this as $U_1(t) = \rho_1 \cdot t + M_1$ where $M_1 = \sup_{T_1 \leq t < T_1 + d_1}(f_1(t) - \rho_1 \cdot t)$;
- define the lower boundary of the pseudo-periodic behavior of $f_2$ as the line $L_2(t)$ such that $\forall t \geq T_2, f_2(t) \geq L_2(t)$. We can compute this as $L_2(t) = \rho_2 \cdot t + m_2$ where $m_2 = \inf_{T_2 \leq t < T_2 + d_2}(f_2(t) - \rho_2 \cdot t)$;
- then, we can bound $\bar{t}$ through $U_1(\bar{t}) \leq L_2(\bar{t})$, i.e., $\bar{t} \geq \frac{M_1 - m_2}{\rho_2 - \rho_1}$. Note that, by construction, this bound is valid only if $\bar{t} \geq \max(T_1, T_2)$.

Then, $f_1 \wedge f_2$ is pseudo-periodic with $T = \max(T_1, T_2, \bar{t})$, $d = d_1$, $c = c_1$.

*Proof:* We distinguish two cases, i.e., when the last intersection between $f_1$ and $f_2$ occurs before or after the start of their pseudo-periodic behaviors.

If before, it is enough to consider $T = \max(T_1, T_2)$ to have all the information for $f_1 \wedge f_2$. Otherwise, we can use the bound $\bar{t} \geq \max(T_1, T_2)$ as $T$ for the same purpose.

Both cases are represented by the expression $T = \max(T_1, T_2, \bar{t})$, which guarantees that $(f_1 \wedge f_2)(t) = f_1(t) \; \forall t \geq T$. Then:

$$\forall t \geq \max(T_1, T_2, \bar{t})$$
$$(f_1 \wedge f_2)(t + d_1) = f_1(t + d_1) \wedge f_2(t + d_1)$$
$$= f_1(t + d_1)$$
$$= f_1(t) + c_1$$
$$= f_1(t) \wedge f_2(t) + c_1.$$

$\square$

The above theorems allow us to compute $T_{f \wedge g}, d_{f \wedge g}$ and $c_{f \wedge g}$ for any pair of operands $f, g$. We will then need to compute the minimum in interval $D_{f \wedge g} = [0, T_{f \wedge g} + d_{f \wedge g}[$. In order to do this, we will need sequences $S_f^{D_f}, S_g^{D_g}$, with $D_f = D_g = D_{f \wedge g}$.

*Convolution*

First, we decompose both functions, $f$ and $g$, in their transient and periodic parts: $f = f_t \wedge f_p$, where
- $f_t(t) = f(t) \; \forall t \in [0, T_f[; f_t(t) = +\infty$ otherwise;
- $f_p(t) = f(t) \; \forall t \in [T_f, +\infty[; f_t(t) = +\infty$ otherwise.

Then, we can decompose the convolution as:

$$f \otimes g = f_t \wedge f_p \otimes g_t \wedge g_p$$
$$= (f_t \otimes g_t) \wedge (f_t \otimes g_p) \wedge (f_p \otimes g_t) \wedge (f_p \otimes g_p).$$

We discuss first how the above partial convolutions are computed, and then what are the properties of the final result.

For the first term, we observe that $f_t \otimes g_t$ is defined in $[0, T_f + T_g[$ and is equal to $+\infty$ for $t \geq T_f + T_g$.

For the second and third terms, we have the following result:

*Theorem 6 (Convolution of Transient and Periodic Part):* $f_t \otimes g_p$ is pseudo-periodic from $T_f + T_g$ with period $d_g$ and increment $c_g$. The symmetric result holds for $f_p \otimes g_t$.

*Proof:* Since $f_t(t) = +\infty$ for all $t \geq T_f$, we can write – for all $t \geq 0$:

$$(f_t \otimes g_p)(t) = \inf_{0 \leq s < T_f}(f_t(s) + g_p(t - s)).$$

Then, for $t \geq T_f + T_g$, $0 \leq s < T_f \implies t - s \geq T_2$. Thus,

$$(f_t \otimes g_p)(t + d_p) = \inf_{0 \leq s < T_f}(f_t(s) + g_p(t + d_p - s))$$
$$= \inf_{0 \leq s < T_f}(f_t(s) + g_p(t - s)) + c_g$$
$$= (f_t \otimes g_p)(t) + c_g.$$

$\square$

Finally, for the fourth and last term:

*Theorem 7 (Convolution of Periodic Parts):* $f_p \otimes g_p$ is pseudo-periodic from $T = T_f + T_g + d$, with length $d = \text{lcm}(d_f, d_g)$ and increment $c = d \cdot \min(\rho_f, \rho_g)$.

To compute the convolution, we need sequences
- $S_{f_p}$ with $D_{f_p} = [T_f, T_f + 2d[$;
- $S_{g_p}$ with $D_{g_p} = [T_g, T_g + 2d[$.

Their convolution $S_{f_p \otimes g_p}$ is then computed over interval:

$$D_{f_p \otimes g_p} = [T_f + T_g, T_f + T_g + 2d[.$$

*Proof:* Since $f_p$ and $g_p$ are defined as above, we can write:

$$(f_p \otimes g_p)(t) = \inf_{0 \leq s \leq t} f_p(s) + g_p(t - s)$$
$$= \inf_{T_f \leq s \leq t - T_g} f_p(s) + g_p(t - s)$$
$$= \inf_{a \geq T_f, b \geq T_g, a + b = t} f_p(a) + g_p(b).$$

Then, for $t \geq T_f + T_g + d$:

$$(f_p \otimes g_p)(t + d) = \inf_{a \geq T_f, b \geq T_g, a + b = t + d} f_p(a) + g_p(b)$$
$$= \left(\inf_{a' \geq T_f, b' \geq T_g, a' + b' = t} f_p(a' + d) + g_p(b')\right) \wedge$$
$$\left(\inf_{a' \geq T_f, b' \geq T_g, a' + b' = t} f_p(a') + g_p(b' + d)\right).$$

Here, we split the infimum in the min of two expressions, where term $d$ appears as an argument of $f_p$ and $g_p$, respectively, so that later on we can leverage the pseudo-periodic property for either function. To a closer inspection, the first case is equivalent to limiting $a \in [T_f + d, t + d], b \in [T_g, t]$, whereas the second term is equivalent to limiting $a \in [T_f, t], b \in [T_g + d, t + d]$. One can verify that this covers all the possible cases, thus the split is valid.

Furthermore, having $a \in [T_f + d, t + d]$ (in the first term) and $b \in [T_g + d, t + d]$ (in the second term) means that we are, for all values in the range, to the right of the first pseudo-period of size $d$ for that function, and we can thus apply the pseudo-periodic property.

Since $d = \text{lcm}(d_f, d_g)$, it is a multiple of both $d_f$ and $d_g$. We can compute $k_f = d/d_f$ and $k_g = d/d_g$, $k_f, k_g \in \mathbb{N}$.

Thus, the two terms can be written as:

$$(f_p \otimes g_p)(t+d) =$$
$$= \left(\inf_{a \geq T_f, b \geq T_g, a+b=t} f_p(a) + g_p(b) + k_f \cdot c_f\right)$$
$$\wedge \left(\inf_{a \geq T_f, b \geq T_g, a+b=t} f_p(a) + g_p(b) + k_g \cdot c_g\right)$$
$$= \inf_{a \geq T_f, b \geq T_g, a+b=t} f_p(a) + g_p(b) + \min(k_f \cdot c_f, k_g \cdot c_g)$$
$$= (f_p \otimes g_p)(t) + \min(k_f \cdot c_f, k_g \cdot c_g)$$
$$= (f_p \otimes g_p)(t) + d \cdot \min(\rho_f, \rho_g).$$

□

*Minimum of the four terms:* It is important to observe that – in the general case – we cannot compute a priori the value of $T$ for the minimum of the four terms. This is relevant, since it is what forces one to implement convolution by decomposing it into the four partial convolutions – plus a minimum – described above.

In fact we have that:
- the first term, $f_t \otimes g_t$, has no pseudo-periodic behavior, so it does not affect this discussion;
- the second term, $C_2 := f_t \otimes g_p$, has $\rho_2 = \rho_g$;
- the third term, $C_3 := f_p \otimes g_t$, has $\rho_3 = \rho_f$;
- the fourth term, $C_4 := f_p \otimes g_p$, has $\rho_4 = \min(\rho_f, \rho_g)$.

Consider the case of $\rho_f > \rho_g$. We can compute $\min(C_2, C_4)$ to have $T = \max(T_2, T_4) = T_f + T_g + d$ and $d = \text{lcm}(d_2, d_4) = \text{lcm}(d_f, d_g)$. However, for $\min(C_3, \min(C_2, C_4))$ we have $T = \max(T_f + T_g + d, \bar{t})$, where $\bar{t}$ is the bound for the intersection of the two curves, which we cannot determine *a priori* as we do not know the shape of the partial convolutions.

This is not true if $\rho_f = \rho_g$, however, in which case we can determine that the minimum of the partial convolution has $T = T_f + T_g + d$. In fact, under this hypothesis, we can compute the entire convolution in a single by-sequence convolution. We thus obtain the following:

*Theorem 8 (Convolution of Curves With the Same Slope):* If $\rho_f = \rho_g (= \rho)$, then $f \otimes g$ has:
- $T = T_f + T_g + d$;
- $d = \text{lcm}(d_f, d_g)$;
- $c = \rho \cdot d$.

The entire convolution can be computed using sequences $S_f^{D_f}, S_g^{D_g}$ with $D_f = [0, T_f + 2d[, D_g = [0, T_g + 2d[$, and then computing the convolution of these sequences, $S_{f \otimes g}^D$, over the interval $D = [0, T_f + T_g + 2d[$.

*Proof:* See the above discussion. □

## APPENDIX B
## FORMAL PROOF FOR PERIOD MINIMIZATION

We prove formally that periods are integer multiples of the minimal period, as discussed informally in Section IV-A1.

*Theorem 9:* Let $f \in \mathcal{U}$ be a non-UA function, and let $\tilde{d}$ be its minimal period. Then, for any period $d$, it holds that

$$d/\tilde{d} \in \mathbb{N}.$$

*Proof:* We define the integer part of $d/\tilde{d}$ as $k := \lfloor d/\tilde{d} \rfloor$ and the fractional part of $d/\tilde{d}$ as $q := d/\tilde{d} - \lfloor d/\tilde{d} \rfloor \in [0, 1[$, thus $d/\tilde{d} = k + q$. Then, it holds that

$$\begin{aligned}
f(t+d) &= f(t + (k+q)\tilde{d}) \\
&= f(t + qd + k \cdot \tilde{d}) \\
&\stackrel{(5)}{=} f(t + q\tilde{d} + (k-1) \cdot \tilde{d}) + c \\
&\stackrel{(5)}{=} \cdots \\
&\stackrel{(5)}{=} f(t + q\tilde{d}) + k \cdot c.
\end{aligned}$$

As the period $\tilde{d}$ is assumed to be minimal, if $0 < q < 1$ then we observe $f(t + q\tilde{d}) = f(t) + q\tilde{c}$ cannot apply, thus the UPP property does not hold for $d$. Thus, $d$ cannot be an equivalent representation unless $q = 0$. □

ACKNOWLEDGMENT

The authors would like to thank Paul Nikolaus of the DISCO Laboratory, Technische Universität Kaiserslautern, for useful discussions on the formal aspects of the article, and the anonymous reviewers, whose suggestions considerably improved this article.

**Raffaele Zippo** is currently pursuing the Ph.D. degree with the Universities of Pisa and Florence. He has coauthored several papers appeared in journals and conferences in his research areas. He is the primary author of Nancy, an open-source Network Calculus library. He has been involved in industrial research projects, and is the coauthor of one patent. His research interests include worst-case analysis of heterogeneous systems and algorithms for NC computations.

**Giovanni Stea** received the Ph.D. degree from the University of Pisa, Italy, in 2003. He is a Full Professor with the Department of Information Engineering, University of Pisa. He has coauthored more than 120 peer-reviewed articles and 17 patents in his research areas. He has been involved in national and EU research projects, and he has led joint research projects with industrial partners. His current research interests include quality of service and resource allocation in networks, performance evaluation, and multi-access edge computing. He was a member of the technical and/or organization committees for several international conferences, including SIGCOMM and VTC. He has served on the Editorial Board of the *Wireless Networks* journal.

Open Access funding provided by 'Università di Pisa' within the CRUI CARE Agreement